# The Effect of Sporadic Dormancy on Adaptation under Natural Selection: A Formal Theory

Author: Sasanka Sekhar Chanda

Affiliation: Indian Institute of Management Indore

Designation: Professor, Strategic Management

Email: sasanka2012@gmail.com

Address: C-101, Academic Block, IIM Indore

Prabandh Shikhar, Rau-Pithampur Road,

Indore, MP, India 453556

Phone number: +917312439591

Date: June 28, 2025.

**Acknowledgement**

I thank Prof. Animangsu Ghatak of the Indian Institute of Technology, Kanpur for encouragement, suggestions and feedback, that helped improve the paper significantly. All errors remain my sole responsibility.

# The Effect of Sporadic Dormancy on Adaptation under Natural Selection: A Formal Theory

**Abstract.** Researchers puzzle over questions as to how rare species survive extinction, and why a significant proportion of microbial taxa are dormant. Computational simulation modeling by a genetic algorithm provides some answers. First, a weak/rare/lowly-adapted species can obtain significantly higher fitness by resorting to sporadic dormancy; thereby the probability of extinction is reduced. Second, the extent of fitness-gain is greater when a higher fraction of the population is dormant; thus, the probability of species survival is greater for higher prevalence of dormancy. In sum, even when the environment is unfavorable initially and remains unchanged, sporadic dormancy enables a weak/rare species enhance the extent of favorable adaptation over time, successfully combating the forces of natural selection.

-----------------------------------------------------------------------

-----------------------------------------------------------------------

Dormancy connotes a reversible state of low metabolic activity in a microorganism, e.g., bacteria, archaea, and microeukaryotes, etc. (Jones and Lennon, 2010; Shoemaker and Lennon, 2018), spanning timescales from instantaneous responses, to seasonal, millennial and even geologic timescales (Bradley, 2025; Cano and Borucki, 1995; Vreeland, Rosenzweig, and Powers, 2000). Dormant individuals do not take part in reproduction (Lennon and Jones, 2011; Shade, 2024; Shoemaker and Lennon, 2018). Dormancy is a very common state of microbial existence in the world (Shade, 2024); perhaps it is the default mode of most bacterial life (Lewis, 2007). At any given time, 90% of the bacteria and fungi in soil are dormant (Barron, 2024). Besides, 20 – 80% of the bacteria recovered from environmental samples appear to be metabolically inactive (Cole, 1999; Jones and Lennon, 2010). The pervasive prevalence of dormancy suggests that "the primary activity of many biological systems is to do nothing" (Melnikov, 2024). This remark, though made in a humorous vein, carries a deeper message: dormancy may have a profound role in species survival.

Dormancy serves as a strategy to persist in harsh conditions and throughout unfavorable changes in their environment (Bradley, 2025; Jones and Lennon, 2010). For example, seed dormancy enables vital processes—like germination, seedling emergence and establishment—to occur when environmental conditions are favorable (Chahtane et al., 2017; Loayza et al., 2023; Shu et al., 2016). Moreover, dormancy allows bacteria to withstand adverse conditions and resume activity when circumstances improve (Stevenson, 1977). Besides, certain microorganisms form spores (a dormant stage) to resist stressors such as temperature, desiccation, and antibiotics (Jones and Lennon, 2010).

Prior research suggests that dormant individuals become members of a seed bank that functions as a storage reservoir; repeated transitions to and from the seed bank helps maintain high levels of microbial biodiversity (Jones and Lennon, 2010). These reservoirs provide "migration from the past" to the current population and influence the current evolutionary response, given that the dormant fraction of the population is not exposed to contemporary selection pressures (Yamamichi et al., 2019). The underlying argument appears to be that in fluctuating environments, dormancy maintains genetic and phenotypic diversity via the storage effect, thereby reducing the probability of local extinctions (Locey et al., 2020).

I offer two points of divergence to complement the viewpoint above. Several species do not invest resources towards the maintenance of network components—receptors, readout molecules, and energy to perform work (Govern and ten Wolde, 2014)—necessary for responsive transitioning between dormant and active states (Lennon et al., 2021). Thus, it is worthwhile to study whether certain species can enhance survival probability through acquisition of favorable adaptations under natural selection by resorting to sporadic dormancy[1] *even when environment is unchanged, say from an initially unfavorable state or from an initially favorable state*. This inquiry is distinct from the phenomenon that environmental change can convert a maladaptation of an earlier time to a favorable adaptation at a later point in time, according some adaptation advantage (Kearns and Shade, 2018). Second, even though waking up from dormancy injects diversity in a species population, it is not clear how or why such additional diversity should enhance favorable adaptation or survival probability. Infusion of diversity does not always enhance fitness (Chanda and Burgelman, 2025). Thus, it is necessary to explain under what conditions *fitness*—or favorable adaptation—is enhanced under sporadic dormancy, and conditions where such improvement does not materialize.

Experimental research about the effects of sporadic dormancy on species survival has some limitations. It is difficult to measure *fitness* accurately (Primack and Kang, 1989). Moreover, it is difficult to predict how evolution will shape the fortunes of a species of microorganisms—since long-term phenotypic and molecular data are scarce, constraining researchers to rely on investigations through space, in lieu of time (Orsini et al., 2013). Besides, in species with complex life-histories, estimating long-term fitness can be challenging (Agrawal et al., 2021). Thus, it is somewhat a tall order to estimate long-term success for species with dormancy on the basis of short-term measures of fitness (Agrawal et al., 2021).

---

[1] The adjective 'sporadic' signifies wide-prevalence of dormancy, and random entry into and exit from dormancy.

Formal theory development by computational simulation modeling—the approach undertaken in this study—can somewhat alleviate above issues. This approach allows studying an idealized model of a phenomenon (McKelvey, 1997), encompassing clearly-delineated interactions by a parsimonious set of variables. The emergence of a macro-level outcome—longer-term fitness attained by a species—can be studied from the ground up, based on micro-level interactions comprising natural selection allowing *only* fitter members to contribute traits that get inherited by progeny. It is also feasible to maintain certain other key assumptions in a prior seminal study (Jones and Lennon, 2010)—bidirectional and repeatable transition between active and dormant states, not requiring passage through the dormant state for microbial reproduction, and consideration of both the active and dormant fractions of the microbial community. Besides, one is in a position to fashion populations as having significant maladaptation (weak species / rare taxa) and other populations having significant extent of favorable adaptation (strong species). This computational modeling platform enables us theorize "how and when dormancy can influence evolution as a population genetic process" (Shoemaker and Lennon, 2018, p. 60).

In particular, we are in a position to investigate an interesting scenario: the effect of natural selection on longer-term fitness of a species that starts off as severely maladapted to the environment (a *weak* species)—and the environment remains unchanged—in absence of dormancy, and under sporadic dormancy. The computational modeling results show that when a significant fraction of a weak species stays dormant, there is large gain in fitness over time, in comparison to a situation of nil-dormancy. This suggests that rare (lowly-endowed/ maladapted) taxa (weak species) can avoid extinction threat merely by taking recourse to sporadic dormancy—regardless of whether such taxa have the wherewithal for environmental sensing and notwithstanding the fact that the unfavorable environment at the beginning remained unchanged. Moreover, the results show that higher extent of fitness gain—and hence higher chances of survival—accrue under higher levels of dormancy. This explains why a large fraction of the biological system appears to be inactive, at any given point in time.

Above results further uphold the view from prior research that dormancy *preserves existing genetic diversity by decreasing the rate that genetic diversity is removed from the population* (Hairston and De Stasio, 1988; Koopmann et al., 2017; Shoemaker and Lennon, 2018; Vitalis, Glémin, and Olivieri, 2004). However, further computational modeling results show that, in an unchanging environment, a *strong* species—favorably adapted to the

environment—obtains *lower* fitness under sporadic dormancy. In this case, dormancy functions as an impediment to fuller utilization of favorable adaptation. This upholds the intuition that preservation of diversity by sporadic dormancy does not lead to higher fitness in all cases.

The computational modeling results for a *changing environment* show that both strong (well-endowed) and weak (lowly-endowed) species stand to obtain higher fitness under sporadic dormancy, in agreement with a large body of extant research.

**MATERIALS AND METHODS**

In order to construct the simulation model, I invoke a genetic algorithm drawing from Holland (1975), March (1991), and Chanda and Miller (2019). Detailed specifications of the model are provided in *Section* **A** of the *Supplementary Document*. In brief, the environment (ENV) is an *M*-bit string, initially populated with random values. The population code is also an *M*-bit string that is empty initially. Each member of the focal species is an *M*-bit string, whose values are randomly populated at first, and then modified to increase maladaptation (to construct a weak species) or favorable adaptation (to construct a strong species). There are *N* members. The task of the algorithm is to get into the population-code as many values matching that in ENV as possible. This is accomplished by having the traits of only the fitter members updating the population-code, for inheriting by progeny. A member does not take part in reproduction if it is tagged *dormant* in any given time-step. The proportion of matching values between the ENV and the population-code is a measure of *fitness*. The values in ENV don't change if the environment is modeled as stable. The values in ENV change with a certain probability in any time-step, when the environment is modeled as turbulent.

**RESULTS**

In **Fig. 1** I show *Fitness* (the extent of favorable adaptation) attained by a *weak* (lowly-endowed / rare) species after 100 time-steps (generations), in absence of dormancy, and under varying extent of sporadic dormancy. The horizontal axis shows the fraction of members dormant. For a given extent of dormancy, I also vary the extent of maladaptation (*Def*) [2]. The bar-graphs inform about the extent of favorable adaptation (*fitness*) when a species—having a certain extent of weakness or maladaptation or deficiency (*Def*)—finds itself in an unfavorable, unchanging environment and uses varying extent of sporadic dormancy to counter forces of natural selection.

---

[2] Please refer to the *materials and methods* section for details regarding the construction of weak and strong species.

**Fig. 1**. Fitness attained by a weak species, in presence and absence of sporadic dormancy, unchanging environment

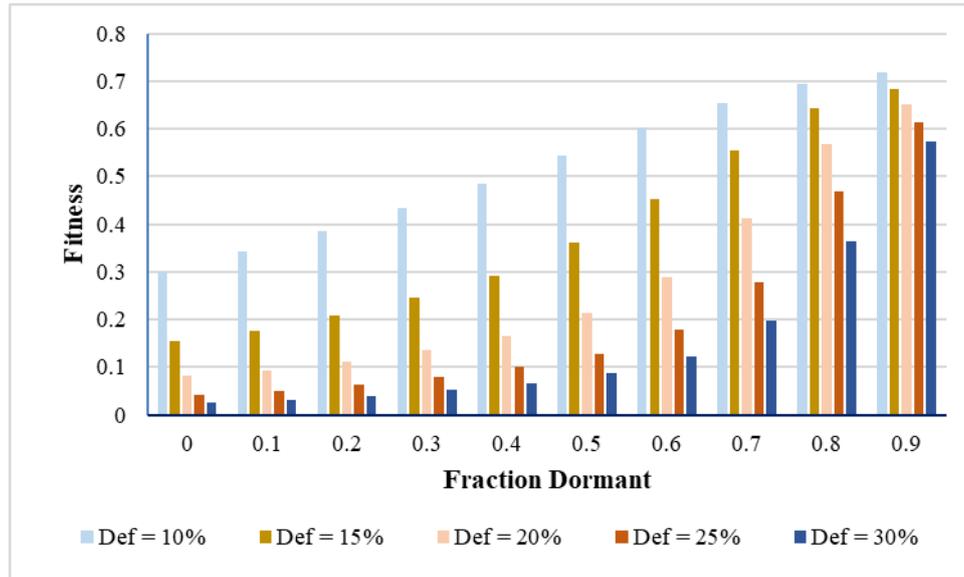

We observe that *Fitness* attained under sporadic dormancy is substantially higher than that in absence of dormancy. For example, we observe that, for ***Def*** = 15% (the second bar from the left, in any cluster) and absent dormancy (***Fraction Dormant*** = 0) *Fitness* attained (15.6%) is much lower than *Fitness* attained (68.4%) under sporadic dormancy (***Fraction Dormant*** = 0.90). Thus, there is a significant increase in the extent of favorable adaptation on account of sporadic dormancy. Hence, we have the following proposition:

> ***Proposition* 1.** In an unchanging environment, for a weak (lowly-endowed/rare) species, sporadic dormancy leads to significant increase in fitness, over time.

Proposition **1** provides an explanation as to how rare/weak species avoid extinction: resorting to sporadic dormancy enables developing a higher level of favorable adaptation, successfully countering the forces of natural selection. In Section **B** of the *Supplementary Document*, I delve into model mechanisms to show how this result materializes. In brief, in the absence of dormancy, a high extent of maladaptation in the member pool seeps into the population-code early on, due to ongoing adaptive efforts comprising the propagation of traits of the fitter members to progeny. Moreover, the diversity in the population drops to zero rather quickly, blocking the avenue for further enhancement of fitness. Thereby, the population settles at a low level of fitness, endangering survival. In contrast, when a significant fraction of the population is dormant, the initial accumulation of maladaptation at the population level is lower, and diversity is preserved for longer. This enables successive generations of progeny attain higher fitness. A higher level of fitness, in turn, enhances survival probability.

We also observe in Figure **1** that higher the fraction of a population that is dormant, greater is the fitness gain on account of sporadic dormancy, for a weak species. This suggests the following proposition:

> ***Proposition* 2.** In an unchanging environment, for a weak (lowly-endowed/rare) species, higher extent of sporadic dormancy leads to higher increase in fitness.

Proposition **2** provides an explanation as to why dormancy is pervasive in microbial taxa. Simply put, pervasive dormancy confers higher survival probability even when an unfavorable environment remains unchanged.

**Fig. 2**. Fitness attained by a strong species, in presence and absence of sporadic dormancy, unchanging environment

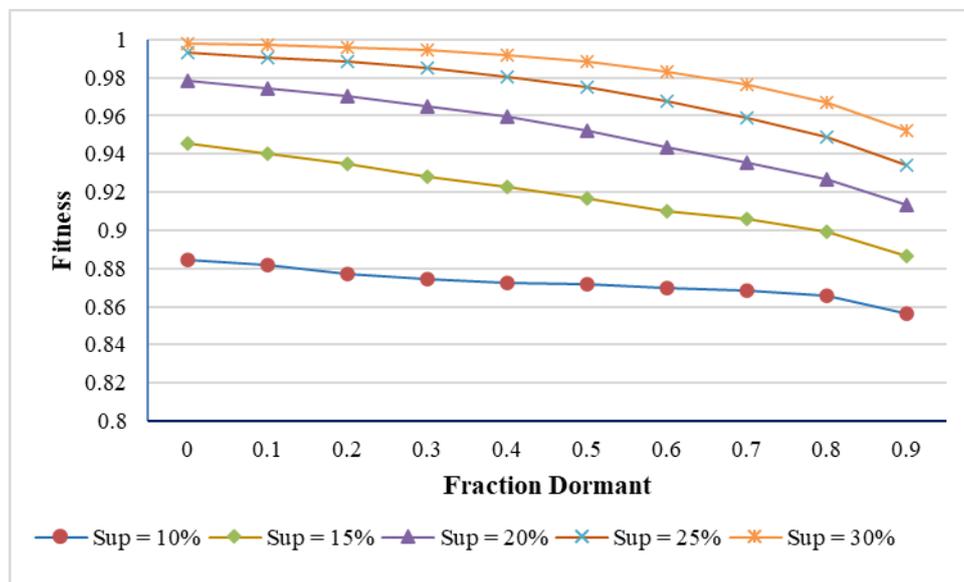

In **Fig. 2** I show *Fitness* attained by a *strong* (well-endowed) species in presence and absence of sporadic dormancy under varying degrees of superiority (*Sup*) of the species population. We observe that in an unchanging environment, sporadic dormancy has a negative effect. For example, for moderate superiority (***Sup*** = 15%), *Fitness* drops from 94% under nil dormancy to 88% under sporadic dormancy, when fraction dormant is 90%. Thus, we have the following proposition:

> ***Proposition* 3.** In an unchanging environment, for a strong species, fitness attained under sporadic dormancy is lower than that attainable when dormancy is absent.

Thus, the impact of sporadic dormancy on species fitness is negative, even though sporadic dormancy helps preserve diversity in the species population. I provide detailed explanation of this result in Section **B** of the *Supplementary Document*. In brief, the extent of favorable adaptation being accumulated in the population-code—under forces of natural selection—is higher when dormancy is absent, because the population comprises members having high extent of favorable adaptations (i.e., since this is a strong species having higher extent of favorable adaptation to the environment, at initiation). For similar reason the extent of accumulation of maladaptation is lower. In contrast, in the presence of sporadic dormancy, accumulation of favorable adaptation is somewhat lower and accumulation of maladaptation is somewhat higher, because far less members are participating, at any given point in time. Thereafter, the nil-dormancy population runs out of diversity and settles at a fairly high level of fitness. The population with sporadic dormancy keeps gaining fitness from the preserved diversity, but is unable to catch up at the time we end simulations (after 100 time-steps).

For a changing (turbulent) environment, further computational simulation modeling shows that both weak and strong species obtain discernible fitness gain by resorting to sporadic dormancy. The simulation results and explanation thereof are provided in Section **C** of the *Supplementary Document*. This result is in agreement with the dominant view in extant research that sporadic dormancy is an apt survival strategy in a changing environment. I present results of additional simulation experiments to reconfirm the propositions under variation of (i) the rate of assimilation of genetic traits of fitter members into the population code and (ii) the rate of migration of population-code traits to progeny in the Section **D** of the *Supplementary Document*.

**DISCUSSION**

The study draws attention to the fortunes of species resorting to sporadic dormancy in the face of forces of natural selection *in an unchanging environment*. For the case where the environment is not favorable to a species—i.e., where the focal species is *weak/rare* on account of having significant extent of maladaptation—we obtain an insight that, merely by resorting to sporadic dormancy, such species can obtain a higher extent of favorable adaptation even if the environment does not turn favorable. This explains how weak/rare species avoid extinction threats. Moreover, higher accrual of fitness gain from higher extent of dormancy—translating into higher survival probability— comes across as an explanation as to why dormancy is pervasive in microbial taxa. For the case where an environment is favorable—i.e., where the focal species is *strong*, on account of having significant extent of favorable adaptation—we learn that, under comparable observation durations, persistence of diversity through sporadic dormancy fails to accord fitness advantage. This complements prior research findings associating persistence of diversity with enhanced fitness and survival probability. Moreover, the dominant view in extant research, that dormancy is a suitable recourse for enhancing fitness *in a changing environment*, is upheld as well.

**Study Limitations**

The study attempts to explain a broad swathe of phenomena using a parsimonious set of variables. It does not explicitly include mechanisms for inter-species rivalry and damage from predation. Rather, members that are fitter (for whatever reason) contribute traits to the population-code that get passed down to progeny. Mutation processes are out of scope. It is also assumed that species members encounter supportive conditions complementing their capabilities for transitioning into and out of dormancy.

**Implications**

The study ties together a range of phenomena pertaining to dormancy, natural selection, adaptation, species rarity and environmental change in one theoretical framework. The theoretical confirmation that sporadic dormancy leads to greater extent of favorable adaptation in rare taxa—*regardless of environmental variation and even in the absence of mutation*—explains why dormancy is vastly prevalent in the biological world and how rare species avoid extinction. It has the potential to narrow down the focus of research attempting to prevent onset of or exit from dormancy in harmful microbes and other research seeking to facilitate transition into and out of dormancy for beneficial microbes. In this regard, experimental research on mechanisms for varying the rate of migration of population-code traits to progeny ($p_1$), and for varying the rate of assimilation of traits of fitter members into the population code ($p_2$) may suggest new intervention techniques. Last but not the least, the computational model can serve as a platform for further theory development—unshackled by limitations of mathematical tractability—clearly tracing micro-to-macro emergence. For example, it is possible to investigate outcomes that materialize over time, when members of a species differ with respect to rate of migration of population-code traits to progeny ($p_1$), or when a subset of members undergoes varying extent of mutation, in stable and changing environments. Moreover, minor modifications can allow researchers study variables of interest like average the amount of time required for a beneficial allele to reach quasi-fixation (Shoemaker and Lennon, 2018), under experimentally modulated timings for entry and exit into dormancy (Ayati and Klapper, 2012) as well as switching responsively vs. stochastically in changing environments (Malik and Smith, 2008).

**REFERENCES**

Agrawal AA, Hastings AP, Maron JL (2021) Evolution and seed dormancy shape plant genotypic structure through a successional cycle. *Proc. Natl Acad. Sci. USA* **118**(34): e2026212118.

Ayati BP, Klapper I (2012) Models of microbial dormancy in biofilms and planktonic cultures. *Communications in Mathematical Sciences*, **10**, 493–511.

Bradley JA (2025) Microbial dormancy as an ecological and biogeochemical regulator on Earth. *Nature Communications* **16**: 3909. [https://www.nature.com/articles/s41467-025-59167-6]

Barron M (2024) Why dormant microbes matter in a changing climate. *American Society for Microbiology*. [https://asm.org/articles/2024/december/why-dormant-microbes-matter-in-a-changing-climate].

Cano RJ, Borucki MK (1995) Revival and identification of bacterial spores in 25- to 40- million- year- old Dominican amber. *Science*, **268**, 1060–1064.

Chahtane H, Kim W, Lopez-Molina L (2017) Primary seed dormancy: A temporally multilayered riddle waiting to be unlocked. *Journal of Experimental Botany*, **68**(4): 857–869.

Chanda SS, Burgelman RA (2025) Diversity can be the basis for continued organizational order: A computational proof of Prigogine's conjecture bridging physical and social systems. *Research Square*. [https://www.researchsquare.com/article/rs-6521388/v1]

Chanda SS, Miller KD (2019) Replicating agent-based models: Revisiting March's exploration-exploitation study. *Strategic Organization*, **17**(4): 425–449.

Cole JJ (1999) Aquatic microbiology for ecosystem scientists: New and recycled paradigms in ecological microbiology. *Ecosystems* (N Y, Print) **2**:215–225.

Govern CC, ten Wolde PR (2014) Optimal resource allocation in cellular sensing systems. *Proc. Natl Acad. Sci. USA* **111**, 17486–17491 (2014).

Hairston Jr NG, De Stasio Jr BT (1988) Rate of evolution slowed by a dormant propagule pool. *Nature*, **336**, 239–242

Holland JH (1975) *Adaptation in Natural and Artificial Systems*. Ann Arbor, MI: University of Michigan Press.

Jones SE, Lennon JT (2010) Dormancy contributes to the maintenance of microbial diversity. *Proc. Natl Acad. Sci. USA*. **107**(13): 5881–5886. [www.pnas.org/cgi/doi/10.1073/pnas.0912765107].

Kearns PJ, Shade A (2018) Trait-based patterns of microbial dynamics in dormancy potential and heterotrophic strategy: Case studies of resource-based and post-press succession. *ISME Journal*. **12**:2575–2581.

Koopmann B, Müller J, Tellier A, Živković D (2017) Fisher- Wright model with deterministic seed bank and selection. *Theoretical Population Biology*, **114**, 29–39.

Lennon JT, Jones SE (2011) Microbial seed banks: the ecological and evolutionary implications of dormancy. *Nature Reviews: Microbiology*. **9**: 119 – 130.

Lennon JT, den Hollander F, Wilke-Berenguer M, Blath J (2021) Principles of seed banks and the emergence of complexity from dormancy. *Nature Communications* **12**, 4807.

Lewis, K. (2007) Persister cells, dormancy, and infectious disease. *Nature Reviews: Microbiology*. **5**, 48–56.

Locey KJ, Muscarella ME, Larsen ML, Bray SR, Jones SE, Lennon JT (2020) Dormancy dampens the microbial distance–decay relationship. *Philosophical Transactions of the Royal Society B*. **375**(1798):20190243

Loayza AP, García-Guzmán P, Carozzi-Figueroa G, Carvajal DE (2023) Dormancy-break and germination requirements for seeds of the threatened Austral papaya (Carica chilensis). *Scientific Reports* **13**(1):17358. [https://doi.org/10.1038/s41598-023-44386-y]

Malik T, Smith HL (2008) Does dormancy increase fitness of bacterial populations in time- varying environments? *Bulletin of Mathematical Biology*, **70**, 1140–1162.

March JG (1991) Exploration and exploitation in organizational learning. *Organization Science* **2**(1):71–87.

McKelvey B (1997) Quasi-Natural Organization Science. *Organization Science*, **8**(4): 352–380

Melnikov S (2024) Comment on dormancy. In Madeline Barron (2024) Why Dormant Microbes Matter in a Changing Climate. *American Society for Microbiology*. [https://asm.org/articles/2024/december/why-dormant-microbes-matter-in-a-changing-climate]

Orsini L, Schwenk K, De Meester L, Colbourne JK, Pfrender ME, Weider LJ (2013) The evolutionary time machine: Using dormant propagules to forecast how populations can adapt to changing environments. *Trends Ecol Evol.* **28**(5):274-82.

Primack RB, Kang H (1989) Measuring fitness and natural selection in wild plant populations. *Annu. Rev. Ecol. Syst*. **20**, 367–396.

Shade A (2024) Comment on dormancy. In Madeline Barron (2024) Why Dormant Microbes Matter in a Changing Climate. *American Society for Microbiology*. [https://asm.org/articles/2024/december/why-dormant-microbes-matter-in-a-changing-climate]

Shoemaker WR, Lennon JT (2018) Evolution with a seed bank: The population genetic consequences of microbial dormancy. *Evolutionary Applications*, **11**(1), 60-75.

Shu K, Liu X-d, Xie Q, Zu-hua He (2016) Two Faces of One Seed: Hormonal Regulation of Dormancy and Germination. *Molecular Plant* **9**(1): 34-45.

Stevenson LH (1977) A case for bacterial dormancy in aquatic systems. *Microb. Ecol*. 4, 127–133.

Vitalis R, Glémin S, Olivieri I (2004) When genes go to sleep: The population genetic consequences of seed dormancy and monocarpic perenniality. *The American Naturalist*, **163**, 295–311.

Vreeland RH, Rosenzweig WD, Powers DW (2000) Isolation of a 250 million- year- old halotolerant bacterium from a primary salt crystal. *Nature*, **407**, 897–900.

Yamamichi M, Hairston NG, Rees M, Ellner SP (2019) Rapid evolution with generation overlap: The double-edged effect of dormancy. *Theor. Ecol.* **12**, 179–195.

SUPPLEMENTARY DOCUMENT FOLLOWS FROM THE NEXT PAGE

## Section A. Materials and Methods

**Simulation Model**

The environment (ENV) is represented by an *M*-bit string[3]. Each bit can have value "+1" or "-1". At the beginning of a simulation experiment, ENV is randomly populated, with values "+1" and "-1" having equal (one-half) probability of appearing[4]. A given simulation experiment runs for *T* time-steps or time-periods or *generations*. If the environment is configured as *unchanging* or 'stable', the values in ENV remain unchanged across all time-steps. A *changing* or turbulent environment is modeled by having any given bit in ENV flip (from "+1" to "-1" or vice-versa) with a certain probability $p_4$.

There are *N* members of the focal species[5]. Each member or *individual* is represented by an *M*-bit string[6]. At the beginning of a simulation experiment, each member-string is populated by "0", "+1" and "-1" with equal (one-third) probability. We refer to this population as a *Marchian* population, in honor of James March who provided the genetic algorithm invoked in this paper in his seminal article in *Organization Science* in 1991, drawing from the work of Holland (1975). A **weak** (lowly-endowed / rare) **species** is fashioned by having a randomly-chosen *Def* percent[7] member-bits of a *Marchian* population updated with values opposite to that in the corresponding bit-position in ENV. For example, if the third member-bit of a given member is chosen for update, it is updated with a value "-1" if the third bit of ENV is "+1" and updated with "+1" value otherwise. On the other hand, a **strong** (well-endowed) **species** is fashioned by having a randomly-chosen *Sup* percent[8] of member-bits of a *Marchian* population updated with values equal to that in the corresponding bit-position in ENV. In the computer program, deficiency and superiority is set by giving positive and negative values respectively to a variable *v_def*. Thus, on average, the ratio of favorably-adapted to maladapted bits[9] is 1:1 in a *Marchian* population, 2:3 in a population having 15% deficiency and 3:2 in a population having 15% superiority.

In the model, the *population-code* is an *M*-bit string. Conceptually, it functions as the repository of outcomes from genetic selection processes. At the beginning of any simulation experiment, all the bits in the population-code are set to zero. Since there at nil matches with the values in ENV, the population-code has nil extent of favorable adaptation (zero fitness) at initiation. The extent of favorable adaptation—measured by the number bit positions (genotypes) the values (traits) in ENV and the population-code match divided by the total number of bits (*M*)—provides a measure of *Fitness*[10]. During the simulation run, values in the population-code get updated, as do the values in the member strings. The values in member-string bits tend to converge to those in the population-code as the simulation progresses.

---

[3] In some literature, the construct "ENV" is designated as the 'objective function' in the 'problem' being solved.
[4] The values '+1' and '-1' have no intrinsic significance. We could use 'X' and 'Y', with no loss of generality.
[5] In some literature, the construct "member" is described by terms like 'phenotype', 'creature', 'organism' etc.
[6] A bit or cell functions as a 'genotype' or 'chromosome', denoting a placeholder for a property of an individual.
[7] In our graphical outputs, we vary *Def* (abbreviation of *deficiency*) from 10% to 30% in steps of five percent.
[8] In our graphical outputs, we vary *Sup* (abbreviation of *superiority*) from 10% to 30% in steps of five percent.
[9] A genotype having a value "0" is excluded from assessment of favorable adaptation or maladaptation.
[10] In empirical studies, fitness is measured differently: fitness is assumed to be directly proportional to the number of progeny (Primack and Kang, 1989).

In each time-step, fitter members—those having higher extent of favorable adaptations than the population-code—contribute to update of values in population-code. Specifically, if the value in a particular bit-position ("+1" or "-1")[11] appears with a majority of *K* among fitter members, the probability that the value in the population-code gets updated in a given time-step is given by [ 1 – (1 – $p_2$)$^K$ ]. Here $p_2$ is a parameter (set by the experimenter) denoting the rate of assimilation of genetic traits of fitter members into the population code. Moreover, the value in a specific bit-position of a member-string gets updated with the corresponding value (if non-zero) in the population-code with a probability $p_1$. Here $p_1$ is a parameter (set by the experimenter) denoting the rate at which the population-code traits get migrated to progeny.

In sum, a dominant trait among the fitter members gets into the population-code through the $p_2$ process and gets copied into subsequent generations through the $p_1$ process; this illustrates natural selection. We note that some extent of maladaptation—i.e., bit values not matching the corresponding value in ENV—will seep into the population code for the cases where a majority of the fitter individuals have maladaptation.

The specifications above conform to that in March (1991), as elaborated in the conceptual replication by Chanda and Miller (2019). This maps to a scenario of 'Nil Dormancy'. The situation where a fraction of the members of the species is dormant is modeled by having a parameter (**Fraction Dormant** *dorm_fr*) set to a value greater than zero and less than unity. In a given time-step, any member can be dormant with probability *dorm_fr*. In a given time-step, a member tagged 'dormant' takes no part in the $p_1$ and $p_2$ processes i.e., does not take part in reproduction.

The following parameter values apply for the simulation experiments: *M* = 30; *N* = 50; *T* = 100; $p_1$ = 0.50; $p_2$ = 0.50; *dorm_fr* varies from 0 to 0.90; *v_def* = 0.10 to 0.30 (lowly-endowed species) and *v_def* = -0.30 to -0.10 (well-endowed species); $p_4$ = 0 (stable environment) and $p_4$ = 0.01 (turbulent environment)[12]. Each simulation experiment is run 10,000 times with stochastically varying inputs; values shown in the graphical results constitute averages across the 10,000 iterations. In additional experiments $p_1$ and $p_2$ are varied, by configuring with values 0.10 and 0.90.

---

[11] The "0" values in member-bits are ignored, i.e., do not play a role in the update of the population code.
[12] For a turbulent environment, one percent of the bits of the environment (ENV) undergo random change in any time-step. This implies that over 100 periods of simulation the entire environment changes, on average.

## Section B. Model-based explanation of the key results

In the section, I delve into model mechanisms in order to develop an understanding of the genesis of the propositions that in an *unchanging environment*, for a weak (lowly-endowed/rare) species, sporadic dormancy leads to significant increase in fitness, and that for a strong species, fitness attained under sporadic dormancy is lower than that attainable when dormancy is absent.

The scope for motivating the model-based explanations is as follows. I use ***Def*** =15% for *weak* species, and ***Sup*** = 15% for *strong* species, and compare fitness attained at nil-dormancy (***Fraction Dormant*** = 0) and with that attained under sporadic-dormancy (***Fraction Dormant*** = 0.90). In **fig. S1**, I provide a graphic showing fitness attained by weak and strong species under nil-dormancy and under sporadic-dormancy, in an unchanging (stable) environment. Thereafter, in fig. **S2**, **S3** and **S4**, I provide the model-based explanation for the results, by observing the development of three variables—FITNESS, MALADAPTATION and DIVERSITY—over the course of 100 time-steps (generations) of simulation. The results for the strong species are thereafter explained by the graphs in fig. **S5**, **S6**, and **S7**.

**fig. S1:** Fitness attained, in presence and absence of sporadic dormancy, in an unchanging environment

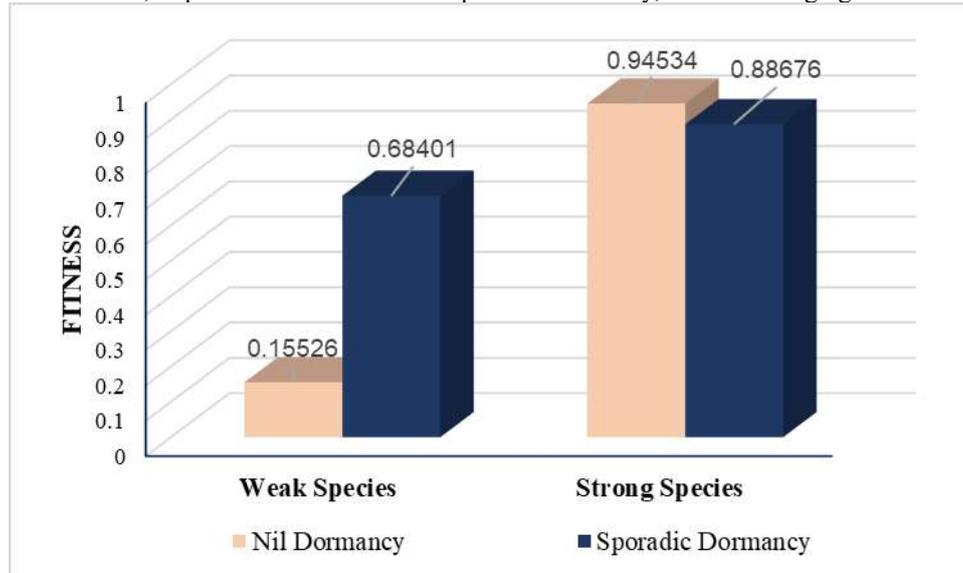

*Parameters.* $p_1$ = 0.50; $p_2$ = 0.50; $p_4$ = 0; $T$ = 100; 10,000 iterations; $v\_def$ = 0.15 and -0.15 for weak and strong species respectively; Dormancy fraction ($dorm\_fr$) = 0 and 0.90 for Nil and Sporadic Dormancy respectively.

**Outcomes for weak species in a stable environment**

In Figure **S2** we note the growth in the population-code *FITNESS* over time, in an *unchanging* (stable) environment corresponding to the graphs for a *lowly-endowed* species in on the lefthand side of Figure **S1** for *nil dormancy* and for *sporadic dormancy*. We observe that when there is nil dormancy, *FITNESS* sharply rises to around ten percent in period two; after about twenty periods of slowing growth, it stabilizes (i.e., does not increase further). In contrast, for a situation involving sporadic dormancy the initial sharp rise in *FITNESS* occurs for longer (till period six) and continues to increase, albeit slowly, and levels off after around time-step 90. This suggests that exclusion of a large section of the (deficient) population leads to higher extent of favorable adaptation. I reason this could be because a large section of the populace having significant extent of maladaptation do not get to contribute to progeny, on account of dormancy.

In order to verify above intuition, in Figure **S3** I present the extent of *MALADAPTATION* in the population-code—computed as the number of nonmatching, non-zero bits between the population-code and ENV. We observe that absent dormancy, *MALADAPTATION* rises to over 85% in the first three periods; but it rises to only about 55% under sporadic dormancy. This suggests that exclusion of a large section of a deficient populace indeed leads to lower extent of maladaptation. We observe that the extent of *MALADAPTATION* decreases over time in both cases. However, for the case involving nil dormancy, this stops by around the 20$^{th}$ time period. This is likely to be the case if diversity has fallen to zero in the species population. We shall check this subsequently. In contrast, the extent of *MALADAPTATION* keeps falling—at slowing rate over time—under sporadic dormancy. This suggests presence of diversity in the population.

The reasoning regarding absence of diversity being responsible for stopping of fitness growth is borne out in Figure **S4**. Diversity contribution from a particular genotype (bit-position or cell) is zero if all members have identical values (traits) for that genotype. I compute the variance of traits for each genotype across population members, average it across the number of bit-positions (*M*) and take square root, to compute the variable *DIVERSITY*. We observe that *DIVERSITY* has indeed fallen to zero around period 20, explaining the freezing of the extent of fitness (and the extent of maladaptation), when dormancy is absent. On the other hand, *DIVERSITY* is well above zero for the case involving sporadic dormancy.

**fig. S2**. Fitness over time, for weak species, unchanging environment

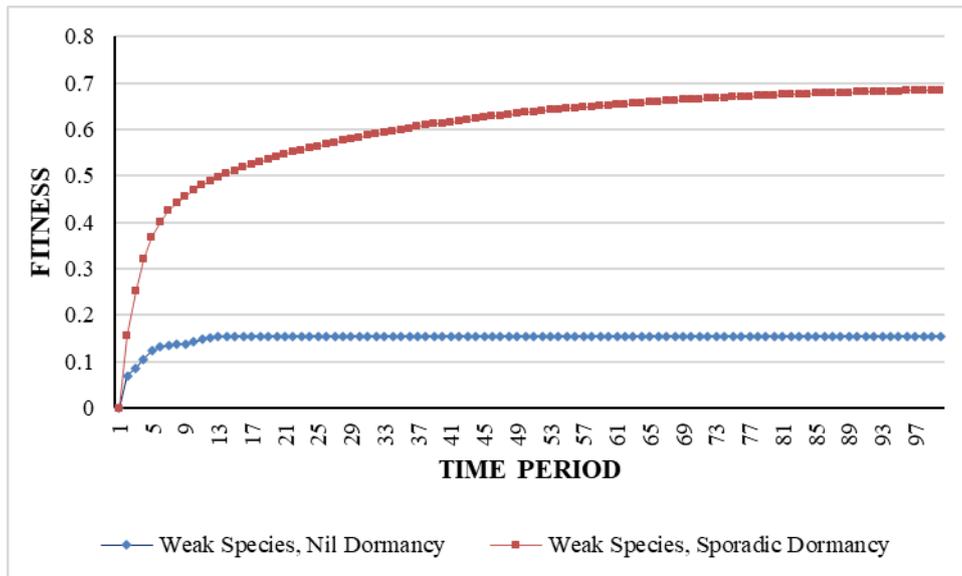

*Parameters*. $p_1$ = 0.50; $p_2$ = 0.50; $p_4$ = 0; *T* = 100; 10,000 iterations; *v_def* = 0.15 for weak species; Dormancy fraction (*dorm_fr*) = 0 and 0.90 for Nil and Sporadic Dormancy respectively.

**fig. S3**. Extent of maladaptation over time, for weak species, unchanging environment

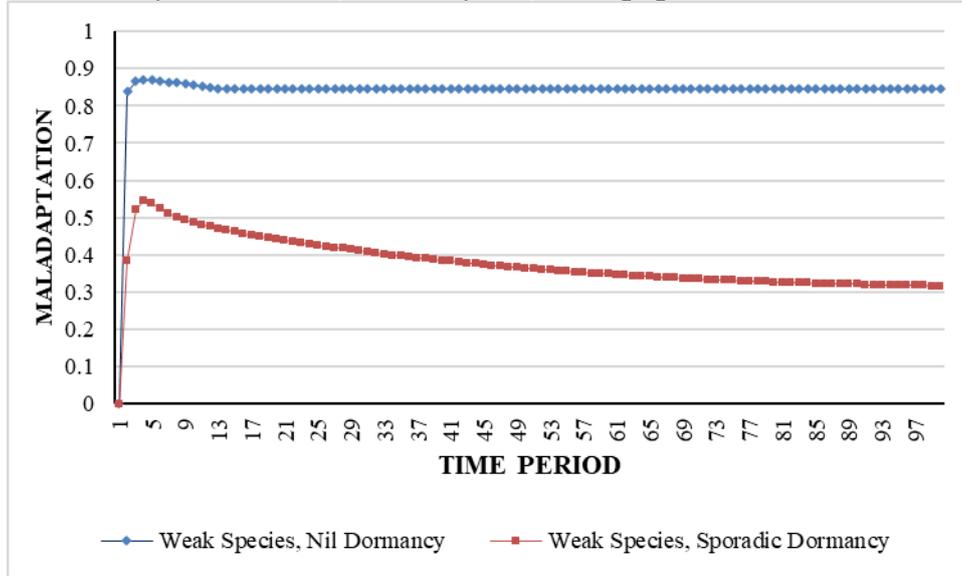

*Parameters.* $p_1 = 0.50$; $p_2 = 0.50$; $p_4 = 0$; $T = 100$; 10,000 iterations; $v\_def = 0.15$ for weak species; Dormancy fraction (*dorm_fr*) = 0 and 0.90 for Nil and Sporadic Dormancy respectively.

**fig. S4**. Diversity in the population over time, for weak species, unchanging environment

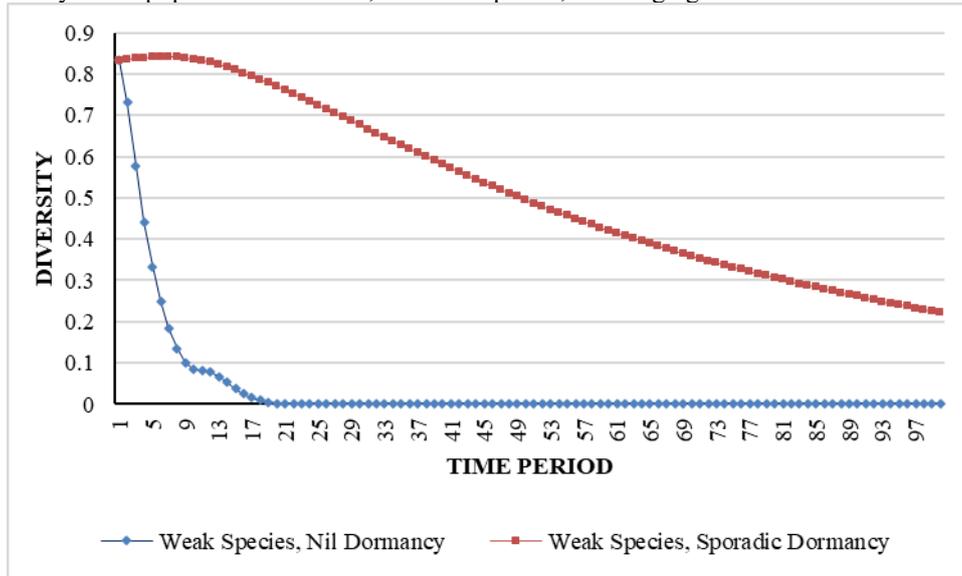

*Parameters.* $p_1 = 0.50$; $p_2 = 0.50$; $p_4 = 0$; $T = 100$; 10,000 iterations; $v\_def = 0.15$ for weak species; Dormancy fraction (*dorm_fr*) = 0 and 0.90 for Nil and Sporadic Dormancy respectively.

The fact that as long as there is diversity in the population, there is scope for maladaptation to decrease and for favorable adaptation to increase can be explained as follows. In the first time-step, fitness of the population-code is zero. Hence all the members who are not dormant are considered fitter than the population-code and get to contribute to update of the population code (by the $p_2$ process). Thereby, the population code gets a certain extent of favorable adaptation and a certain extent of maladaptation. The non-dormant members do not inherit anything from the population-code (by the $p_1$ process), since all values are "0" in the population-code, in the first period.

In the second period, since the population-code has a certain extent of favorable adaptation, non-dormant members in that period face a higher bar for qualifying as 'fitter' than the population-code.

Thus, only a subset of the non-dormant members qualifies, and the dominant traits among them—again in some measure favorable, and in some measure maladapted—update the population-code (by the $p_2$ process). Further, all the non-dormant members also get to inherit the population-code's characteristics (by the $p_1$ process) and pass it to their progeny in the next generation; thereby, some members gain fitness. In the third period, non-dormant members face an even higher bar for qualifying as 'fitter' than the population code. Thereby, valuable traits of the fitter minority get migrated to the population-code. All the while, the non-dormant members keep inheriting the traits accumulating in the population-code (by the $p_1$ process), and pass it to their progeny. This allows a given member to gain fitness; when such fitness exceeds that of the population-code, the member's valuable favorable adaptations get utilized. The process stops when all members have traits similar to those in the population-code. In effect, sporadic dormancy maintains diversity for longer, allowing the species attain overall higher fitness.

**Figure S5**. Fitness over time, for strong species, unchanging environment

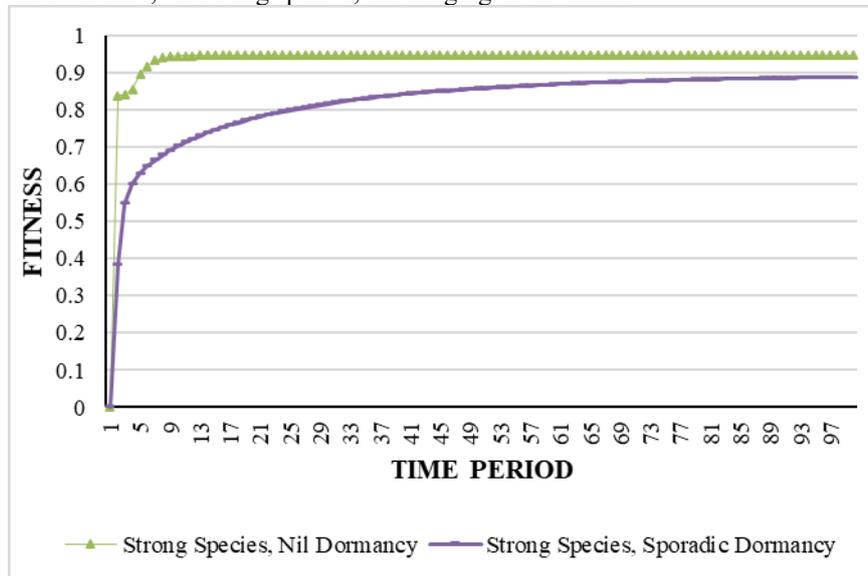

*Parameters.* $p_1 = 0.50$; $p_2 = 0.50$; $p_4 = 0$; $T = 100$; 10,000 iterations; $v\_def = -0.15$ for strong species; Dormancy fraction (*dorm_fr*) = 0 and 0.90 for Nil and Sporadic Dormancy respectively.

**Outcomes for strong species in an unchanging environment**

In **Figure S5** we observe the growth in favorable adaptation (*FITNESS*) over time in an unchanging (*stable*) environment, corresponding to the graphs for a well-endowed species on the righthand side of Figure S2, for *nil-dormancy* and for *sporadic-dormancy*. We observe that, within the first three time-steps, the curve corresponding to nil dormancy attains a *FITNESS* of over 80%; in the same period, the curve for sporadic dormancy attains about 55% *FITNESS*. I reason that fitness is higher in the former case because the population comprises of a large number of members having superior adaptation. The fitness is lower in the latter case because, due to sporadic dormancy, a much leaner set of superior-adaptation members inform the population-code. The picture regarding the extent of maladaptation is expected to be the other way round. This is confirmed in **Figure S6**, where we compare the extent of *MALADAPTATION* over time, for the cases involving nil dormancy and sporadic dormancy. We observe that the extent of *MALADAPTATION* seeping into the population code in the first three periods is around seven percent for nil dormancy, and around 20% for sporadic dormancy. We also observe in Figure **S6** that while the decrease of *MALADAPTATION* flattens around period 20 for the case involving nil dormancy, it continues over the entire observation period for sporadic dormancy. Similar flattening and continuation of trend for the nil-dormancy and the sporadic-dormancy cases are noted in Figure **S5** for *FITNESS* as well. Again, the reason for this is confirmed from **Figure S7**, that shows *DIVERSITY* in the population over time. For the nil-dormancy case, *DIVERSITY* falls to zero around period 20, forestalling any further change in fitness and maladaptation. For the case of sporadic dormancy, presence of *DIVERSITY* ensures increase in fitness and decrease in maladaptation for all time-steps (generations).

**Figure S6**. Extent of maladaptation over time, for strong species, unchanging environment

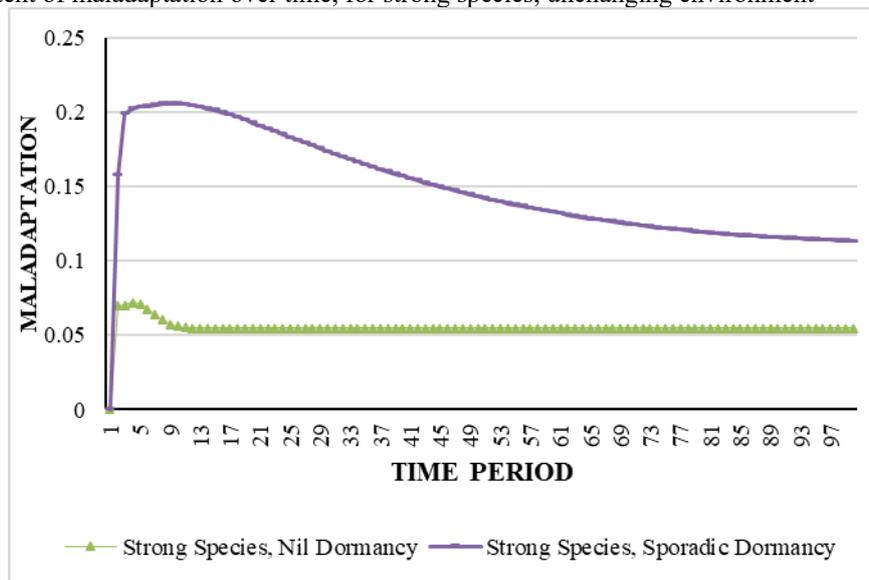

*Parameters*. $p_1$ = 0.50; $p_2$ = 0.50; $p_4$ = 0; $T$ = 100; 10,000 iterations; $v\_def$ = -0.15 for strong species; Dormancy fraction ($dorm\_fr$) = 0 and 0.90 for Nil and Sporadic Dormancy respectively.

**Figure S7**. Diversity in the population over time, for strong species, unchanging environment

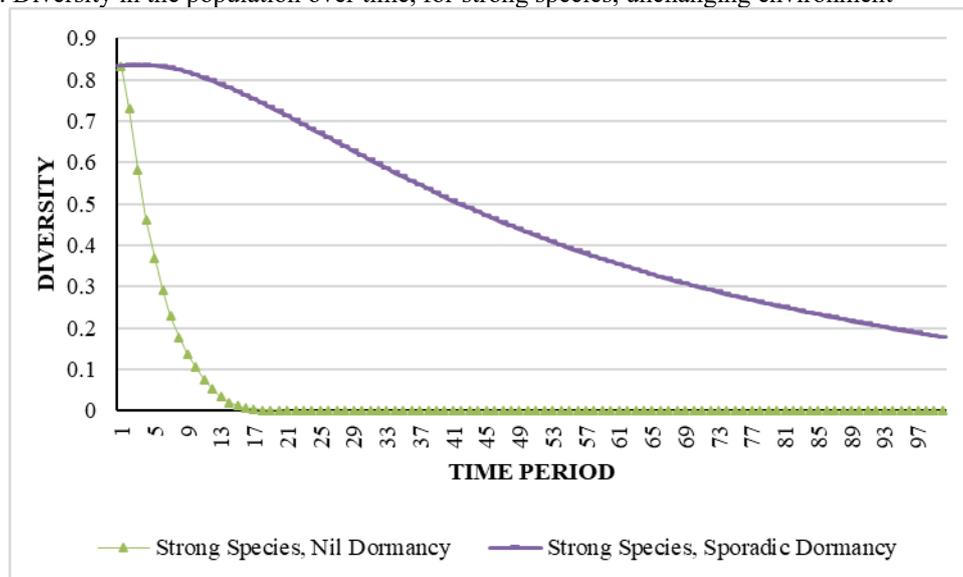

*Parameters*. $p_1$ = 0.50; $p_2$ = 0.50; $p_4$ = 0; $T$ = 100; 10,000 iterations; $v\_def$ = -0.15 for strong species; Dormancy fraction ($dorm\_fr$) = 0 and 0.90 for Nil and Sporadic Dormancy respectively.

## Section C. Results for a changing (turbulent) environment

In the main document, I mention in the passing that in a *changing* (turbulent) *environment*, the simulation modeling results are in agreement with the dominant view in extant research that sporadic dormancy leads to gain of fitness for both strong and weak species. In the section, I first present the graphical results for a *changing environment*—analogous to the results shown in the main document for an unchanging environment. Thereafter, I delve into model mechanisms in order to develop an understanding of the genesis of the results.

The scope for motivating the model-based explanations is as follows. First, in **fig. S8** and **S9**, I provide results for weak and strong species respectively, over varying extent of dormancy and deficiency and superiority in a *changing* (turbulent) *environment*. Thereafter, I use ***Def*** =15% for *weak* species, and ***Sup*** = 15% for *strong* species, and compare *FITNESS* between nil-dormancy (***Fraction Dormant*** = 0) and sporadic-dormancy (***Fraction Dormant*** = 0.90) to motivate model-based explanations of the results for a changing (turbulent) environment, along lines similar to that in Section **B**.

**fig. S8**. Fitness attained by a weak species, in presence and absence of sporadic dormancy, turbulent environment

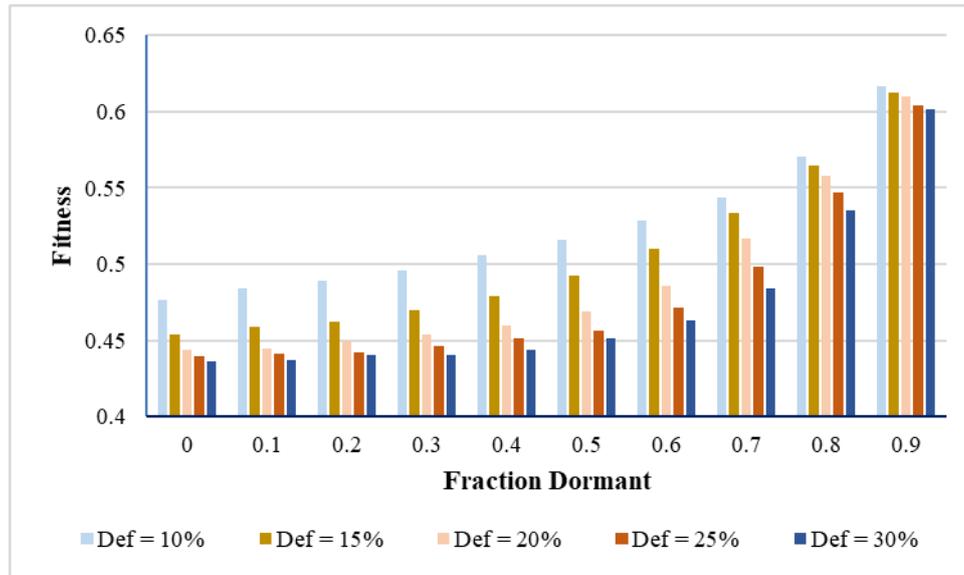

*Parameters*. $p_1$ = 0.50; $p_2$ = 0.50; $p_4$ = 0.01; $T$ = 100; 10,000 iterations; $v\_def$ = 0.15 for weak species; Dormancy fraction (*dorm_fr*) = 0 and 0.90 for Nil and Sporadic Dormancy respectively.

**fig. S9**. Fitness attained by a strong species, in presence and absence of sporadic dormancy, turbulent environment

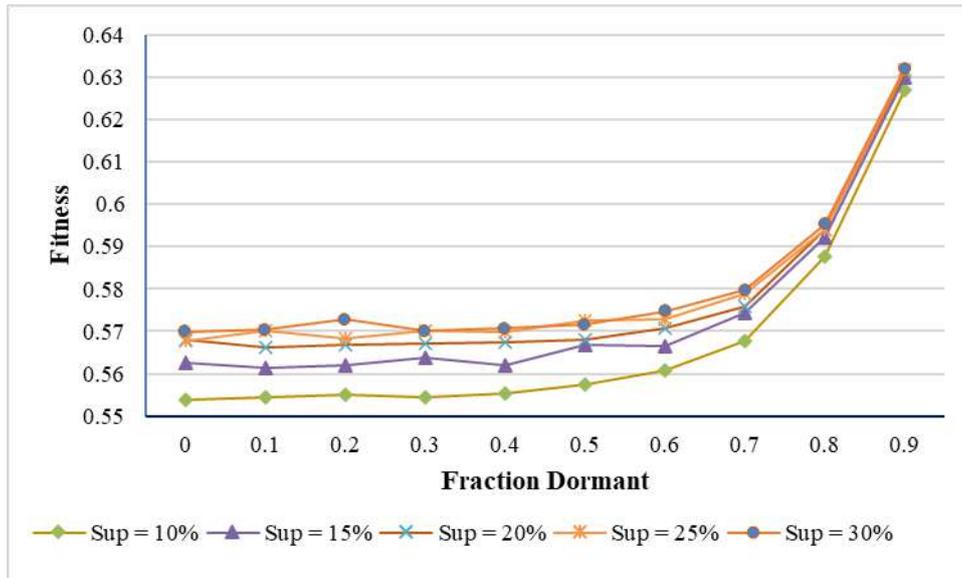

*Parameters.* $p_1 = 0.50$; $p_2 = 0.50$; $p_4 = 0.01$; $T = 100$; 10,000 iterations; $v\_def = -0.15$ for strong species; Dormancy fraction ($dorm\_fr$) = 0 and 0.90 for Nil and Sporadic Dormancy respectively.

**fig. S10:** Fitness attained in presence and absence of sporadic dormancy, in a changing (turbulent) environment

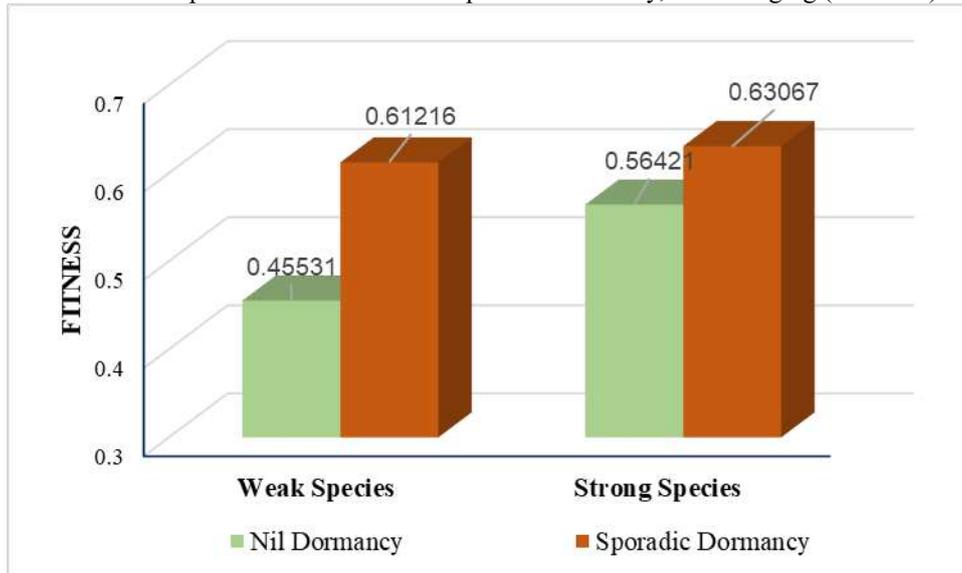

*Parameters.* $p_1 = 0.50$; $p_2 = 0.50$; $p_4 = 0.01$; $T = 100$; 10,000 iterations; $v\_def = 0.15$ and $-0.15$ for weak and strong species respectively; Dormancy fraction ($dorm\_fr$) = 0 and 0.90 for Nil and Sporadic Dormancy respectively.

In **fig. S10**, I provide a graphic showing fitness attained by weak and strong species under nil-dormancy and under sporadic-dormancy, in a changing (turbulent) environment. Thereafter, in fig. **S11**, **S12** and **S13**, I provide the model-based explanation for the results for the weak species, by observing the development of three variables—FITNESS, MALADAPTATION and DIVERSITY—over the course of 100 time-steps (generations) of simulation. The results for the strong species are thereafter explained by invoking the graphics in fig. **S14**, **S15**, and **S16**.

**Outcomes for weak species in a turbulent environment**

In **fig. S11** we observe the growth in favorable adaptation (*FITNESS*) over time in a changing (*turbulent*) environment—corresponding to the graphs for a *lowly-endowed* species on the lefthand side of fig. **S10**, for *nil dormancy* and for *sporadic dormancy*. We observe the curves to be somewhat similar to those in fig. **S2**. However, two important points of difference need to be noted and explained. First, the curve for *nil dormancy* attains a much higher value (45% in **S11** vs. 15% in **S2**). Second, the curve for *sporadic dormancy* attains a somewhat lower value (61% in **S11** vs. 68% in **S2**). These differences occur on account of continuous change due to environmental turbulence. When a bit of ENV flips, the probability that a maladaptation becomes a favorable adaptation is higher if fitness of the population-code is less than 50%. This explains why the nil-dormancy situation attains higher fitness (45% in place of 15%). However, for the sporadic-dormancy case, the population-code fitness is higher than 50%. In this case, an ENV bit flipping on account of turbulence is more likely to convert a favorable adaptation into a maladaptation. This explains why population-code fitness is lower (61% in place of 68%).

**fig. S11**. Fitness over time for weak species, turbulent environment

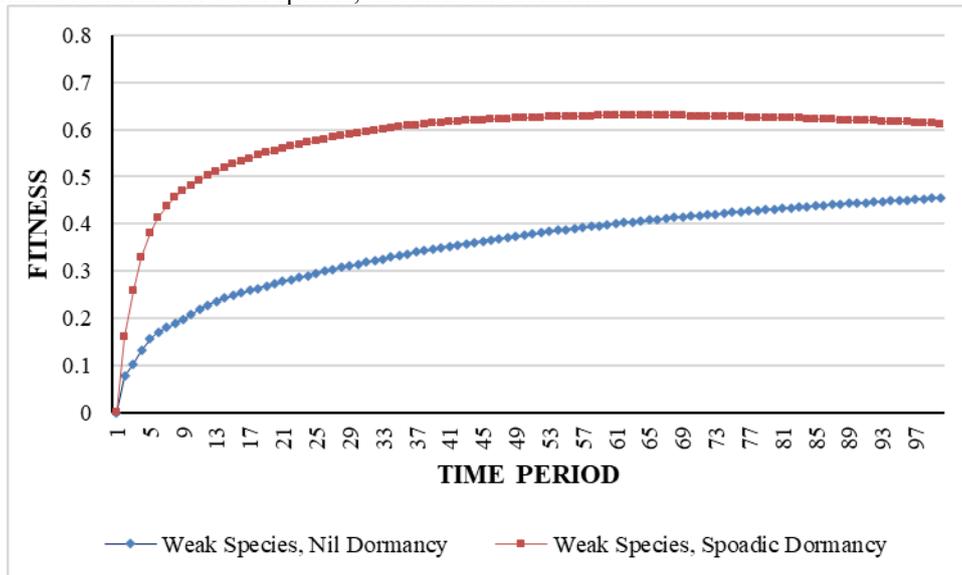

*Parameters*. $p_1 = 0.50$; $p_2 = 0.50$; $p_4 = 0.01$; $T = 100$; 10,000 iterations; $v\_def = 0.15$ for weak species; Dormancy fraction (*dorm_fr*) = 0 and 0.90 for Nil and Sporadic Dormancy respectively.

In **fig. S12** we observe *MALADAPTATION* over time for weak species, in presence and absence of sporadic dormancy, in a *turbulent* environment. The curves are similar to the *unchanging* environment case (fig. **S3**), but the fall in *MALADAPTATION* is steeper in fig. **S12**. As can be confirmed from **fig. S13**, for the nil dormancy case, *DIVERSITY* falls off to zero around period 20. This suggests that, after around period 20, the fall in *MALADAPTATION* for the nil-dormancy case in fig. **S12** is solely due to maladaptation transforming into favorable adaptation on account of environmental turbulence. For the sporadic-dormancy case in fig. **S12**, the enrichment of the population-code from member diversity—lowering *MALADAPTATION*—is somewhat offset by the environmental turbulence tending to increase the extent of maladaptation.

**fig. S12**. Extent of maladaptation over time, for weak species, turbulent environment

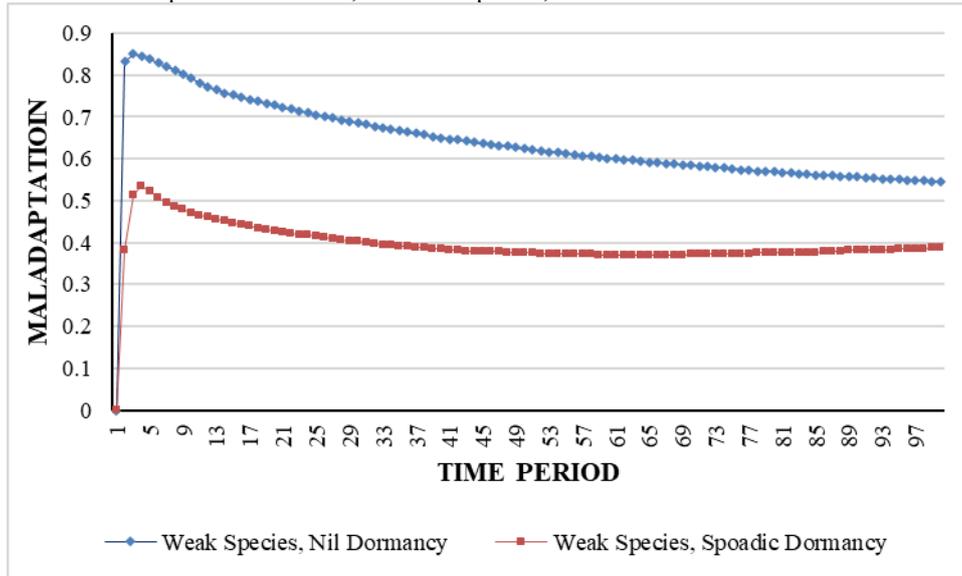

*Parameters.* $p_1$ = 0.50; $p_2$ = 0.50; $p_4$ = 0.01; $T$ = 100; 10,000 iterations; $v\_def$ = 0.15 for weak species; Dormancy fraction (*dorm_fr*) = 0 and 0.90 for Nil and Sporadic Dormancy respectively.

**fig. S13**. Diversity in the population over time, for weak species, turbulent environment

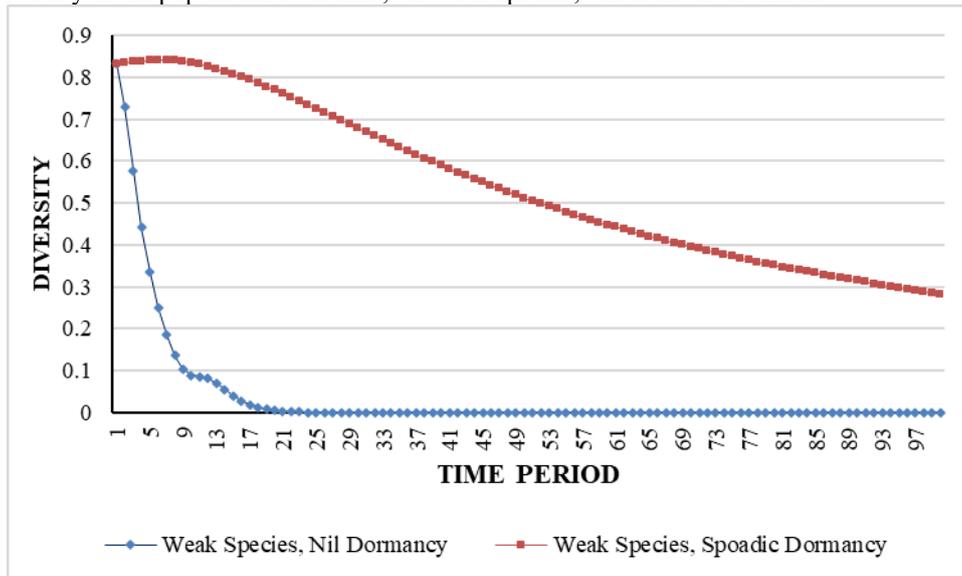

*Parameters.* $p_1$ = 0.50; $p_2$ = 0.50; $p_4$ = 0.01; $T$ = 100; 10,000 iterations; $v\_def$ = 0.15 for weak species; Dormancy fraction (*dorm_fr*) = 0 and 0.90 for Nil and Sporadic Dormancy respectively.

**Outcomes for strong species in a turbulent environment**

In **fig. S14** we observe the growth in favorable adaptation (*FITNESS*) over time in a *turbulent* environment—corresponding to the graphs for a well-endowed species on the righthand side of fig. **S10**, for *nil-dormancy* and for *sporadic-dormancy*. We observe that, as before, by period three, while the nil-dormancy curve attains *FITNESS* of over 80%, the curve for sporadic dormancy attains only about 55% *FITNESS*. Both curves rise for a while thereafter, and then head downwards. In this latter stage, the fall for the nil-dormancy curve is sharper. This is because after period 20, the *DIVERSITY* in the species population falls to zero (**fig. S15**), forestalling the possibility of any further growth in fitness, but environmental turbulence continues to degrade fitness. For the sporadic-dormancy curve the fall is gentler because some of the malevolent effect of environmental turbulence is set-off by fitness increases accruing from continuing *DIVERSITY* in the species population (fig. **S15**). The graphs for *MALADAPTATION* in **fig. S16** confirm this view. Due to environmental turbulence—and owing to the fact that population-code fitness is greater than 50% where environmental turbulence will, on average, turn favorable adaptations into maladaptation— *MALADAPTATION* rises for both the nil-dormancy and sporadic-dormancy cases. However, it is lower for the latter case because some maladaptation is offset by the population-code getting enriched from continuing diversity in the species population.

**fig. S14**. Fitness over time for a strong species, turbulent environment

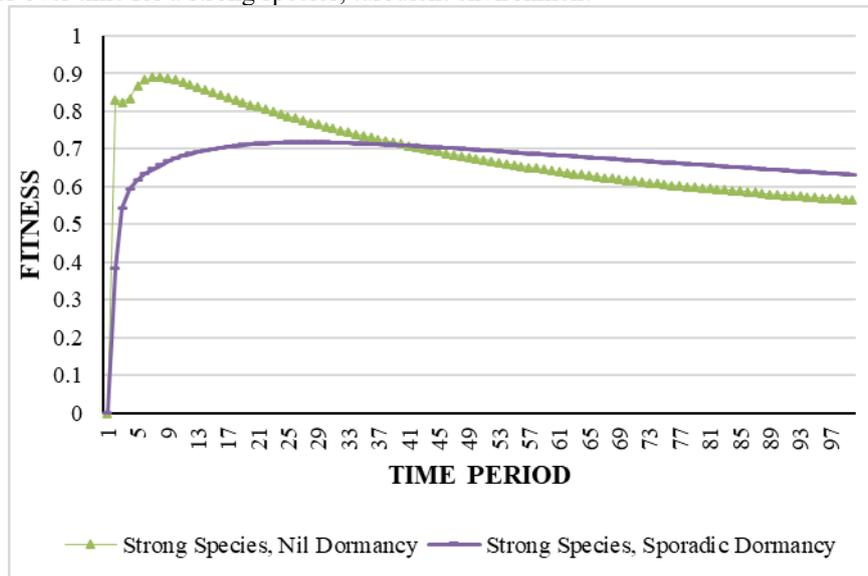

*Parameters*. $p_1$ = 0.50; $p_2$ = 0.50; $p_4$ = 0.01; $T$ = 100; 10,000 iterations; $v\_def$ = -0.15 for strong species; Dormancy fraction ($dorm\_fr$) = 0 and 0.90 for Nil and Sporadic Dormancy respectively.

**fig. S15**. Diversity in the population over time for strong species, turbulent environment

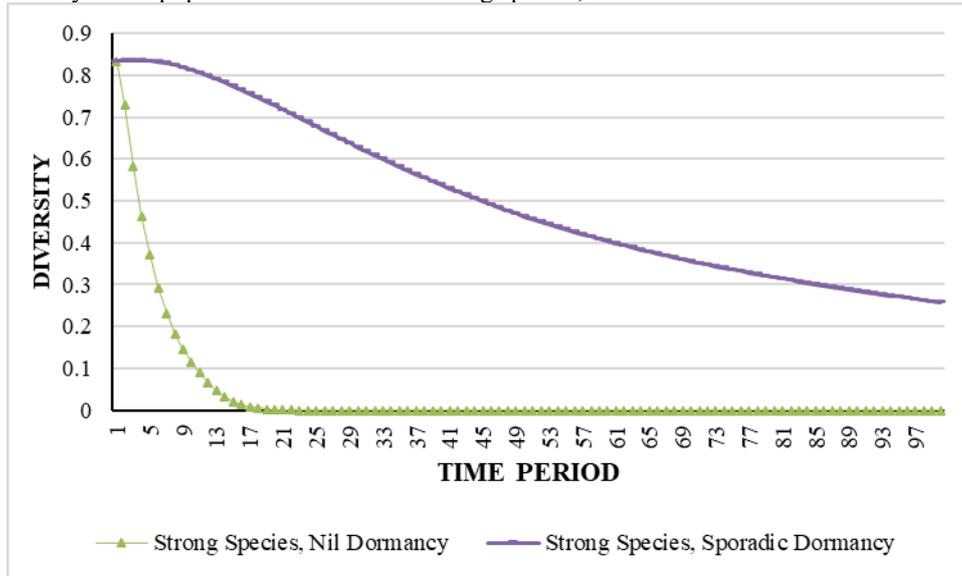

*Parameters.* $p_1 = 0.50$; $p_2 = 0.50$; $p_4 = 0.01$; $T = 100$; 10,000 iterations; $v\_def = -0.15$ for strong species; Dormancy fraction ($dorm\_fr$) = 0 and 0.90 for Nil and Sporadic Dormancy respectively.

**fig. S16**. Extent of maladaptation over time for strong species, turbulent environment

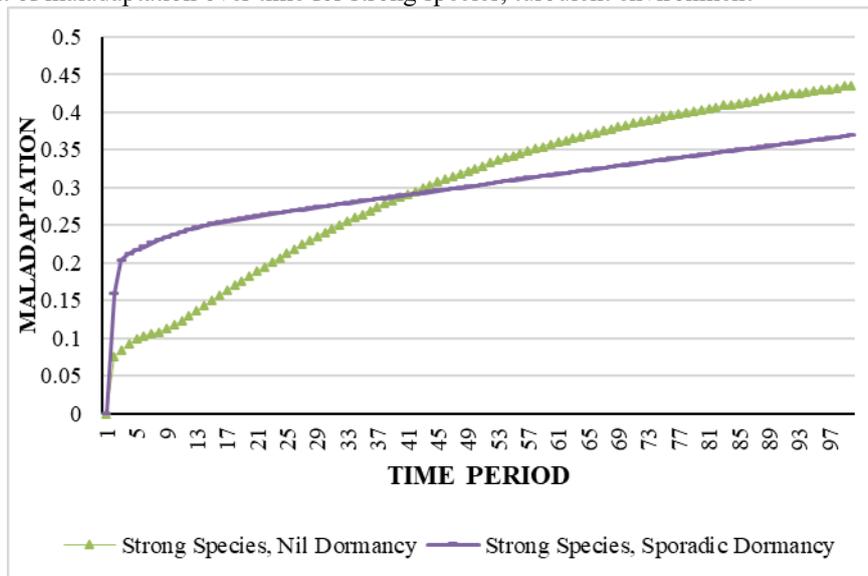

*Parameters.* $p_1 = 0.50$; $p_2 = 0.50$; $p_4 = 0.01$; $T = 100$; 10,000 iterations; $v\_def = -0.15$ for strong species; Dormancy fraction ($dorm\_fr$) = 0 and 0.90 for Nil and Sporadic Dormancy respectively.

## Section D. Variation of the intensity of dormancy, the extent of deficiency and superiority of species, the rate of migration of population-code traits to progeny and the rate of assimilation of genetic traits of fitter members into the population-code

In Figure **1** and Figure **2** in the main document, I have presented the simulation modeling results obtained by varying the extent of deficiency (*v_def*) and the fraction of dormant members (*dorm_fr*) and keeping the values of *p₁* and *p₂* unchanged (at 0.50) for an unchanging (stable) environment for a weak and strong species respectively. In Section **C** of this document, I have presented the corresponding results for a changing (turbulent) environment in fig. **S8** and **S9**, respectively. In this document I present additional results obtained when higher (0.90) and lower (0.10) values of *p₁* and *p₂* are used, keeping other parameters the same. Table **S1** provides a summary.

**Table S1**. Summary of results

| Species | Environment | |
|---|---|---|
| | *Stable* | *Turbulent* |
| *Weak* | $p_1$ = 0.50; $p_2$ = 0.50: Figure **1** | $p_1$ = 0.50; $p_2$ = 0.50: **fig. S8** |
| | $p_1$ = 0.10; $p_2$ = 0.50: **fig. S17** | $p_1$ = 0.10; $p_2$ = 0.50: **fig. S18** |
| | $p_1$ = 0.90; $p_2$ = 0.50: **fig. S21** | $p_1$ = 0.90; $p_2$ = 0.50: **fig. S22** |
| | $p_1$ = 0.50; $p_2$ = 0.10: **fig. S25** | $p_1$ = 0.50; $p_2$ = 0.10: **fig. S26** |
| | $p_1$ = 0.50; $p_2$ = 0.90: **fig. S29** | $p_1$ = 0.50; $p_2$ = 0.90: **fig. S30** |
| *Strong* | $p_1$ = 0.50; $p_2$ = 0.50: Figure **2** | $p_1$ = 0.50; $p_2$ = 0.50: **fig. S9** |
| | $p_1$ = 0.10; $p_2$ = 0.50: **fig. S19** | $p_1$ = 0.10; $p_2$ = 0.50: **fig. S20** |
| | $p_1$ = 0.90; $p_2$ = 0.50: **fig. S23** | $p_1$ = 0.90; $p_2$ = 0.50: **fig. S24** |
| | $p_1$ = 0.50; $p_2$ = 0.10: **fig. S27** | $p_1$ = 0.50; $p_2$ = 0.10: **fig. S28** |
| | $p_1$ = 0.50; $p_2$ = 0.90: **fig. S31** | $p_1$ = 0.50; $p_2$ = 0.90: **fig. S32** |

From the graphs above, I further derive a set of summary results (fig. **33** … fig. **36**) by focusing on the instances where weak and strong species have respectively 15% deficiency and 15% superiority, and comparing nil and sporadic dormancy situations to compute fitness gain (loss) by deducting the fitness obtained under nil-dormancy (***Fraction Dormant* = 0**) from that obtained under sporadic dormancy (***Fraction Dormant* = 0.90**). We observe that the propositions claimed in this study hold good for a range of variation of the rate of migration of population-code traits to progeny **(*p₁*)** and the rate of assimilation of genetic traits of fitter members into the population-code **(*p₂*)**.

Section B1. Results for low values of $p_1$ ($p_1$ = 0.10)

**fig. S17**. Fitness attained by a weak species, unchanging (stable) environment

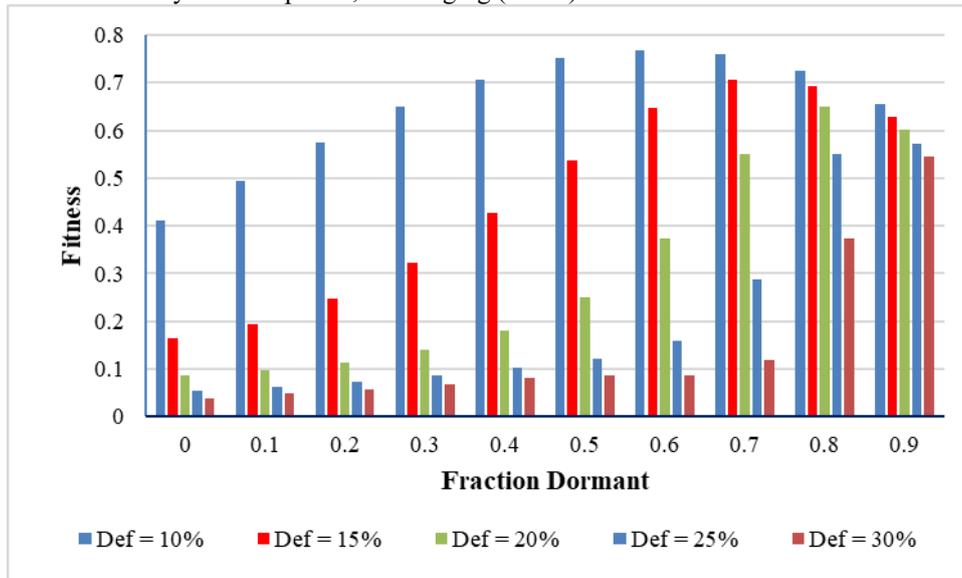

*Parameters*. $p_1$ = 0.10; $p_2$ = 0.50; $p_4$ = 0; $T$ = 100; 10,000 iterations; $v\_def$ = 0.15 for weak species; Dormancy fraction ($dorm\_fr$) = 0 and 0.90 for Nil and Sporadic Dormancy respectively.

**fig. S18**. Fitness attained by a weak species, changing (turbulent) environment

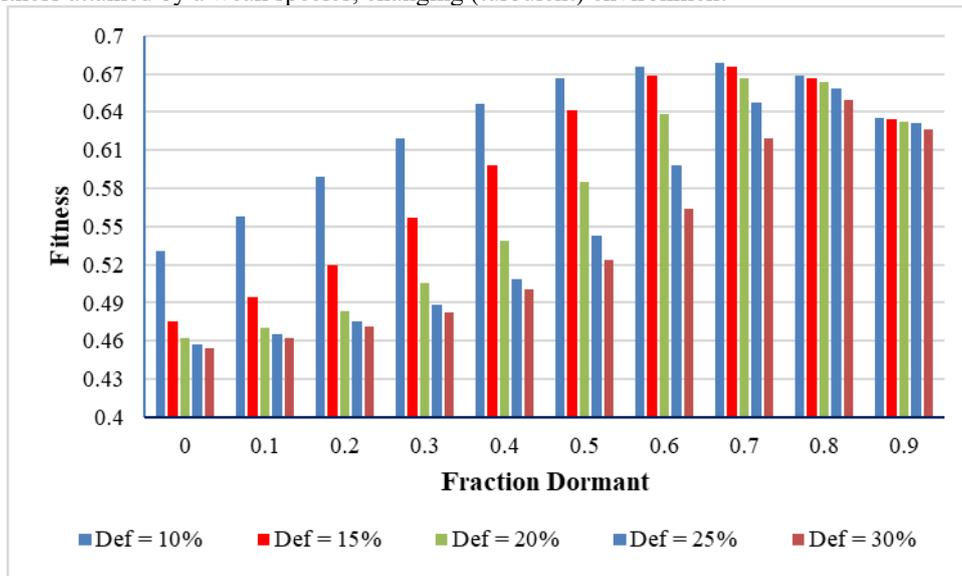

*Parameters*. $p_1$ = 0.10; $p_2$ = 0.50; $p_4$ = 0.01; $T$ = 100; 10,000 iterations; $v\_def$ = 0.15 for weak species; Dormancy fraction ($dorm\_fr$) = 0 and 0.90 for Nil and Sporadic Dormancy respectively.

**fig. S19**. Fitness attained by a strong species, unchanging (stable) environment

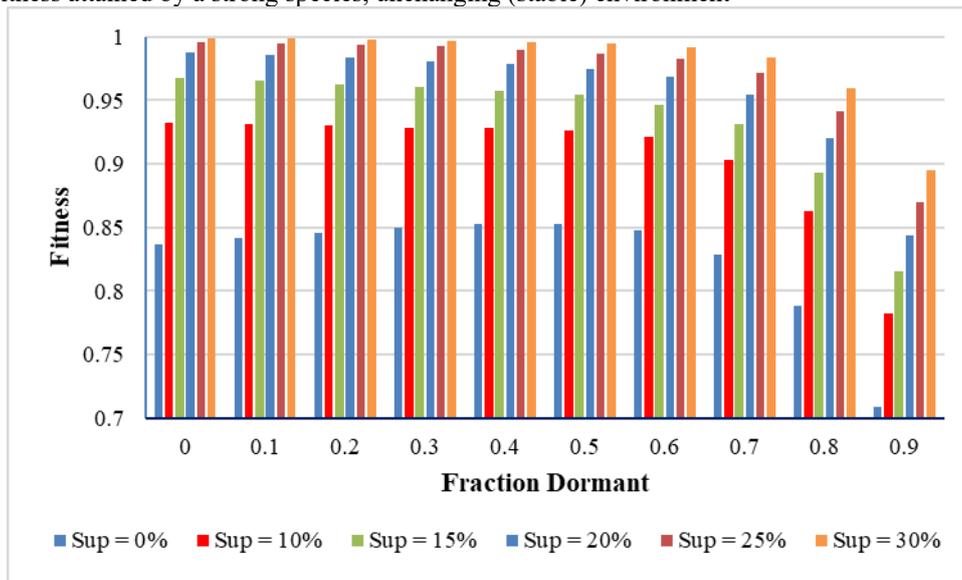

*Parameters.* $p_1$ = 0.10; $p_2$ = 0.50; $p_4$ = 0; $T$ = 100; 10,000 iterations; $v\_def$ = -0.15 for strong species; Dormancy fraction ($dorm\_fr$) = 0 and 0.90 for Nil and Sporadic Dormancy respectively.

**fig. S20**. Fitness attained by a strong species, changing (turbulent) environment

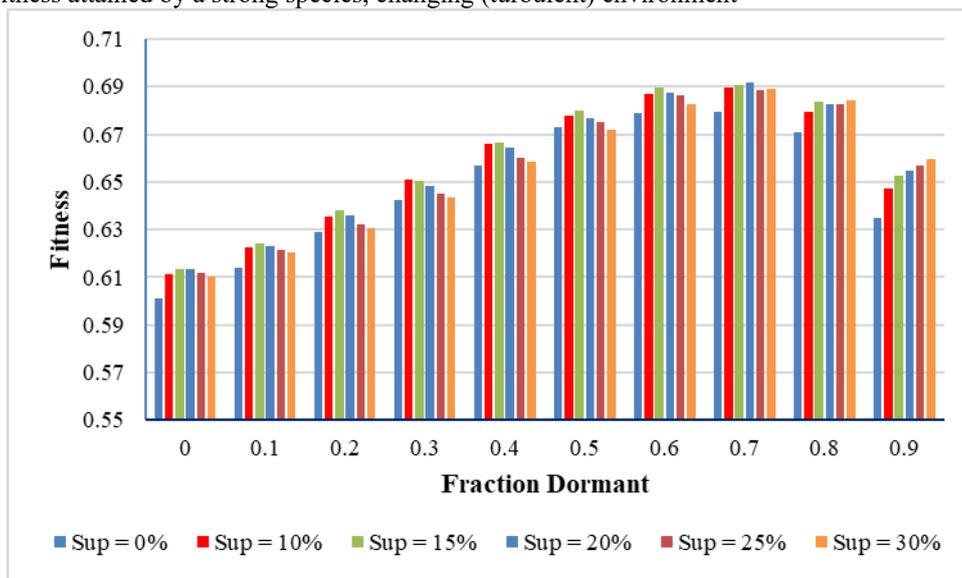

*Parameters.* $p_1$ = 0.10; $p_2$ = 0.50; $p_4$ = 0.01; $T$ = 100; 10,000 iterations; $v\_def$ = -0.15 for strong species; Dormancy fraction ($dorm\_fr$) = 0 and 0.90 for Nil and Sporadic Dormancy respectively.

Section B2. Results for high values of $p_1$ ($p_1$ = 0.90)

**fig. S21**. Fitness attained by a weak species, unchanging (stable) environment

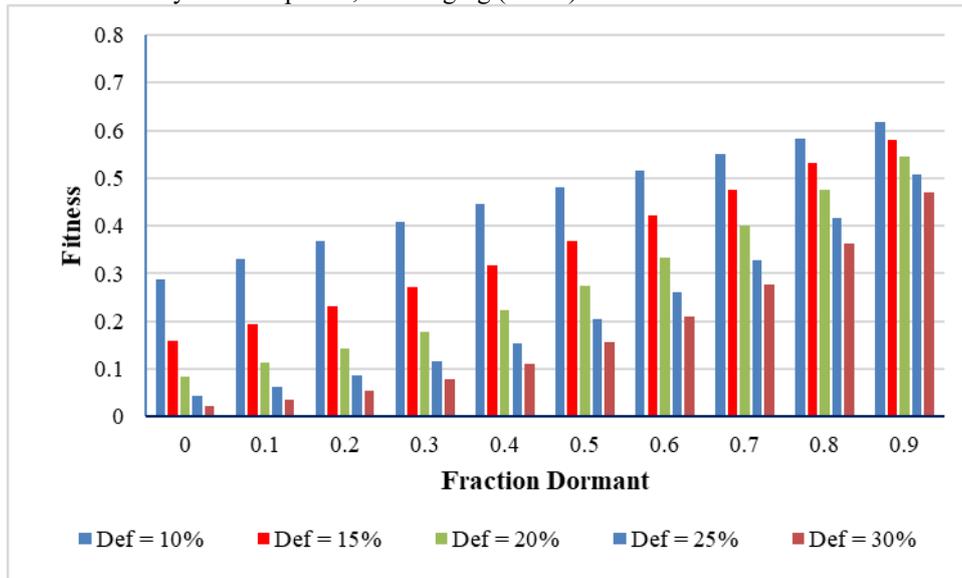

*Parameters.* $p_1$ = 0.90; $p_2$ = 0.50; $p_4$ = 0; $T$ = 100; 10,000 iterations; $v\_def$ = 0.15 for weak species; Dormancy fraction (*dorm_fr*) = 0 and 0.90 for Nil and Sporadic Dormancy respectively.

**fig. S22**. Fitness attained by a weak species, changing (turbulent) environment

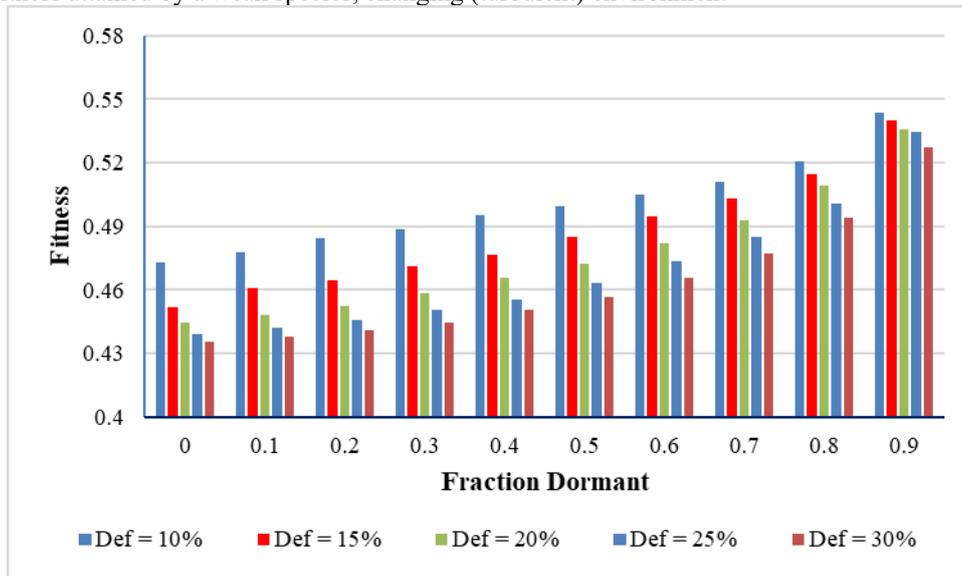

*Parameters.* $p_1$ = 0.90; $p_2$ = 0.50; $p_4$ = 0.01; $T$ = 100; 10,000 iterations; $v\_def$ = 0.15 for weak species; Dormancy fraction (*dorm_fr*) = 0 and 0.90 for Nil and Sporadic Dormancy respectively.

**fig. S23**. Fitness attained by a strong species, unchanging (stable) environment

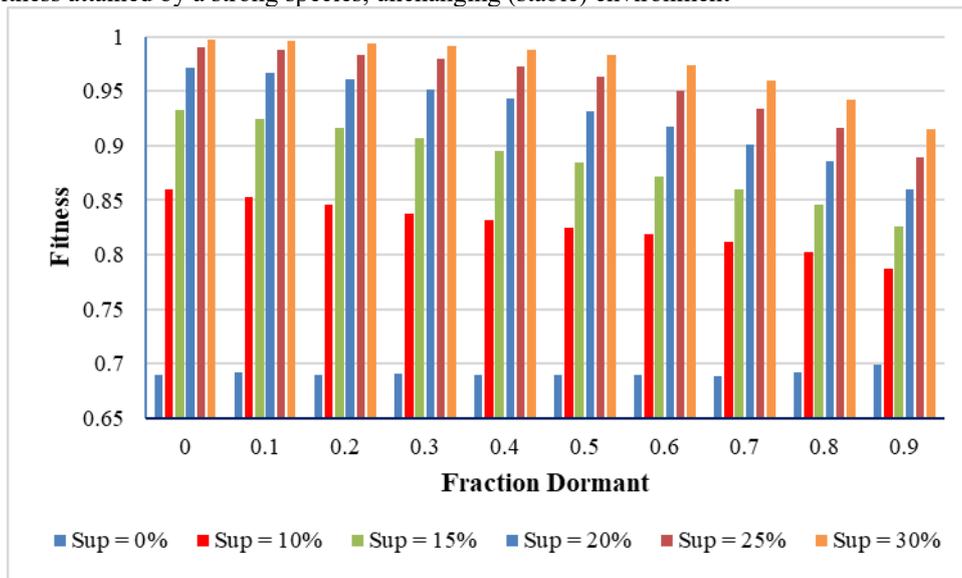

*Parameters.* $p_1 = 0.90$; $p_2 = 0.50$; $p_4 = 0$; $T = 100$; 10,000 iterations; $v\_def = -0.15$ for strong species; Dormancy fraction ($dorm\_fr$) = 0 and 0.90 for Nil and Sporadic Dormancy respectively.

**fig. S24**. Fitness attained by a strong species, changing (turbulent) environment

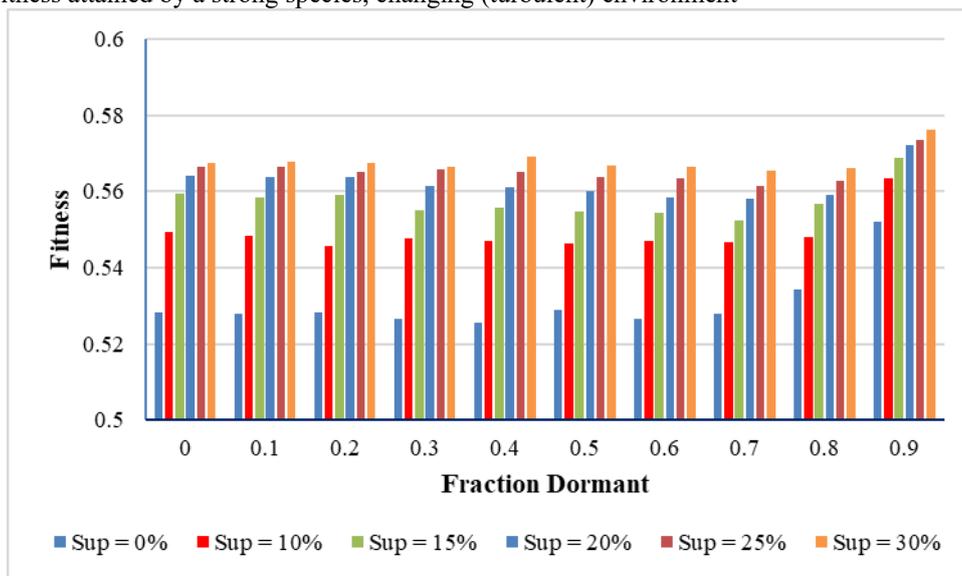

*Parameters.* $p_1 = 0.90$; $p_2 = 0.50$; $p_4 = 0.01$; $T = 100$; 10,000 iterations; $v\_def = -0.15$ for strong species; Dormancy fraction ($dorm\_fr$) = 0 and 0.90 for Nil and Sporadic Dormancy respectively.

Section B3. Results for low values of $p_2$ ($p_2 = 0.10$)

**fig. S25**. Fitness attained by a weak species, unchanging (stable) environment

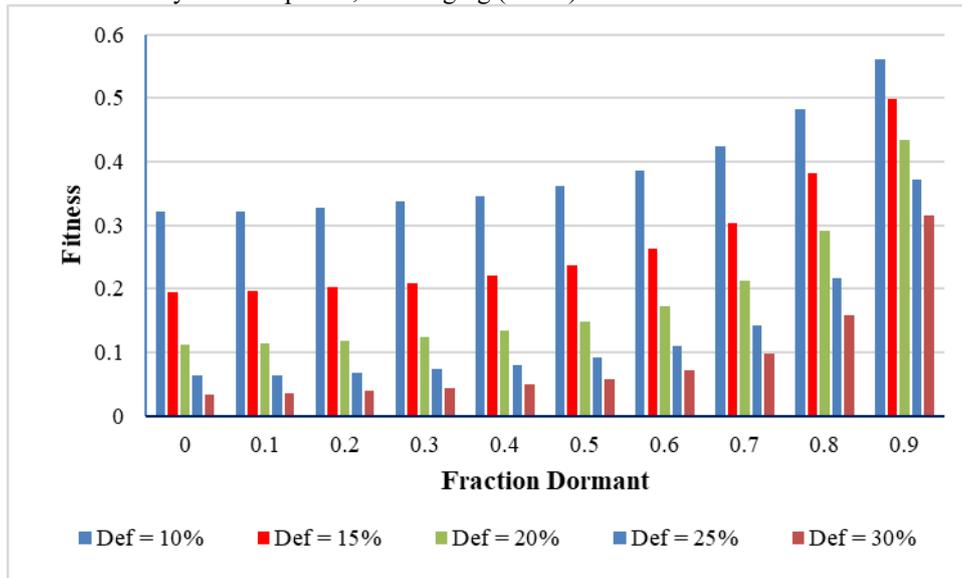

*Parameters*. $p_1 = 0.50$; $p_2 = 0.10$; $p_4 = 0$; $T = 100$; 10,000 iterations; $v\_def = 0.15$ for weak species; Dormancy fraction (*dorm_fr*) = 0 and 0.90 for Nil and Sporadic Dormancy respectively.

**fig. S26**. Fitness attained by a weak species, changing (turbulent) environment

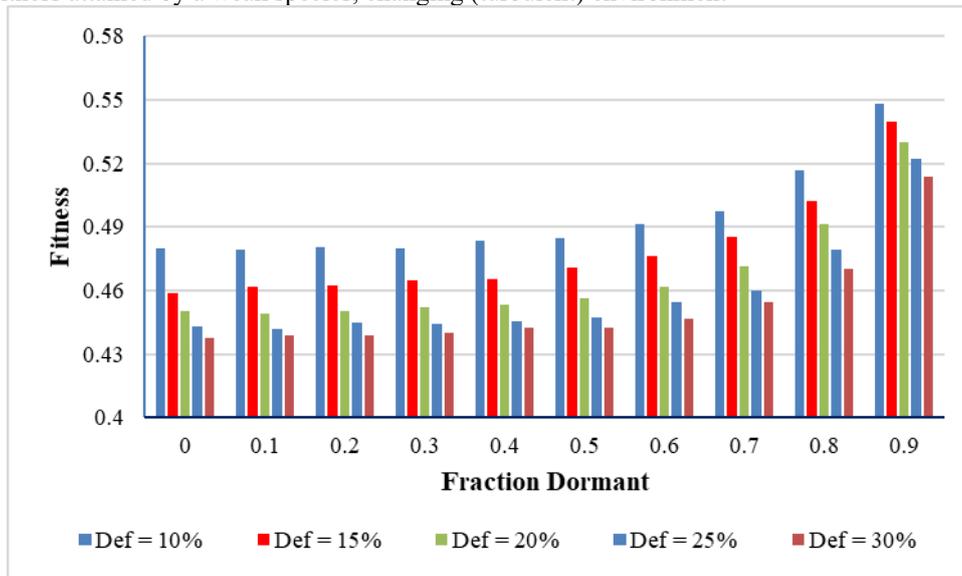

*Parameters*. $p_1 = 0.50$; $p_2 = 0.10$; $p_4 = 0.01$; $T = 100$; 10,000 iterations; $v\_def = 0.15$ for weak species; Dormancy fraction (*dorm_fr*) = 0 and 0.90 for Nil and Sporadic Dormancy respectively.

**fig. S27**. Fitness attained by a strong species, unchanging (stable) environment

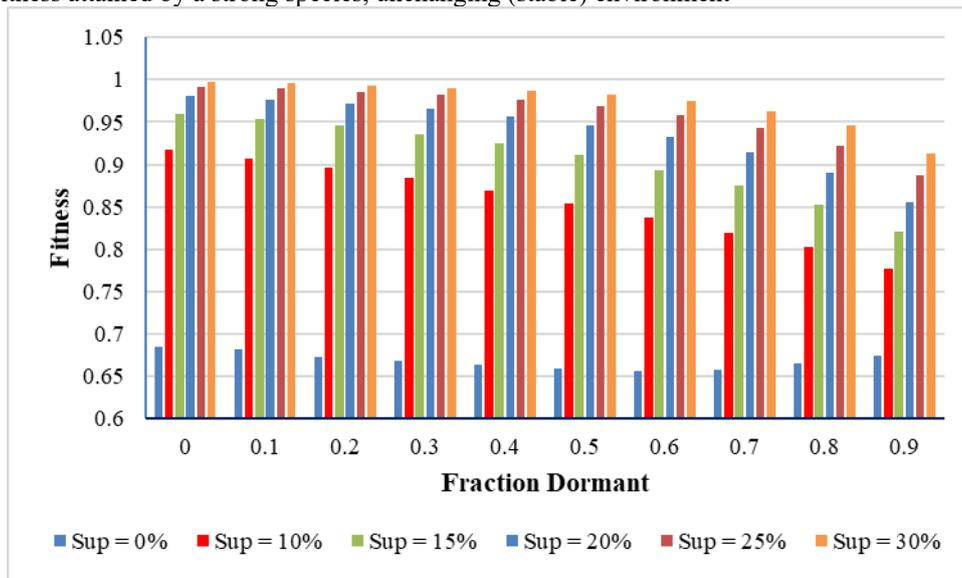

*Parameters.* $p_1 = 0.50$; $p_2 = 0.10$; $p_4 = 0$; $T = 100$; 10,000 iterations; $v\_def = -0.15$ for strong species; Dormancy fraction (*dorm_fr*) = 0 and 0.90 for Nil and Sporadic Dormancy respectively.

**fig. S28**. Fitness attained by a strong species, unchanging (stable) environment

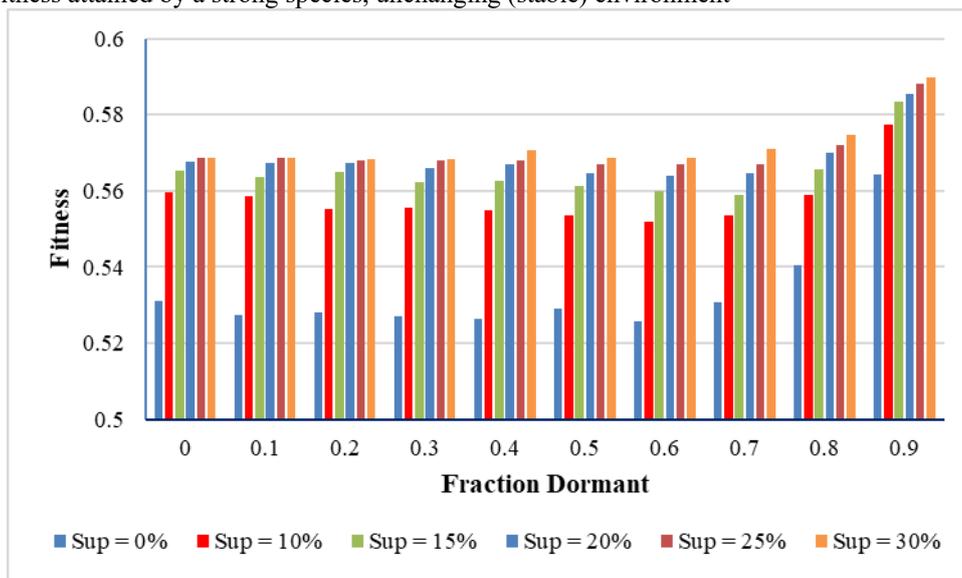

*Parameters.* $p_1 = 0.50$; $p_2 = 0.10$; $p_4 = 0.01$; $T = 100$; 10,000 iterations; $v\_def = -0.15$ for strong species; Dormancy fraction (*dorm_fr*) = 0 and 0.90 for Nil and Sporadic Dormancy respectively.

*Species survival under sporadic dormancy*Section B4. Results for high values of $p_2$ ($p_2 = 0.90$)

**fig. S29**. Fitness attained by a weak species, unchanging (stable) environment

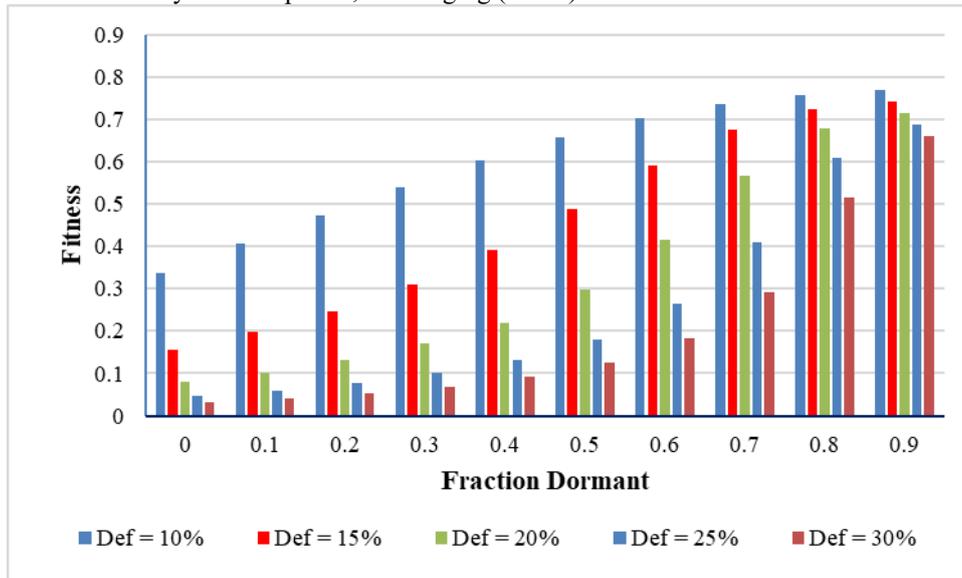

*Parameters*. $p_1 = 0.50$; $p_2 = 0.90$; $p_4 = 0$; $T = 100$; 10,000 iterations; $v\_def = 0.15$ for weak species; Dormancy fraction ($dorm\_fr$) = 0 and 0.90 for Nil and Sporadic Dormancy respectively.

**fig. S30**. Fitness attained by a weak species, changing (turbulent) environment

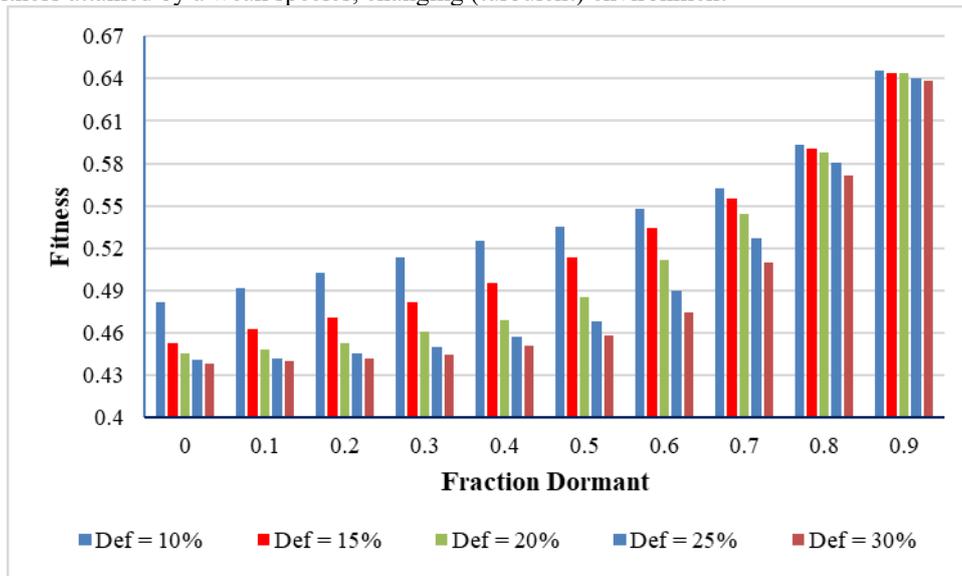

*Parameters*. $p_1 = 0.50$; $p_2 = 0.90$; $p_4 = 0.01$; $T = 100$; 10,000 iterations; $v\_def = 0.15$ for weak species; Dormancy fraction ($dorm\_fr$) = 0 and 0.90 for Nil and Sporadic Dormancy respectively.

**fig. S31**. Fitness attained by a strong species, unchanging (stable) environment

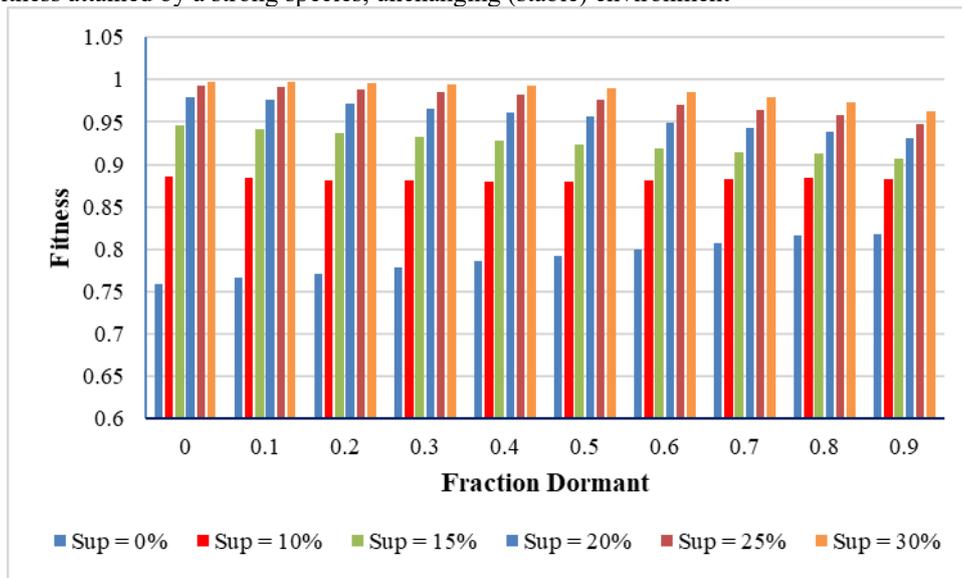

*Parameters*. $p_1$ = 0.50; $p_2$ = 0.90; $p_4$ = 0; $T$ = 100; 10,000 iterations; $v\_def$ = -0.15 for strong species; Dormancy fraction (*dorm_fr*) = 0 and 0.90 for Nil and Sporadic Dormancy respectively.

**fig. S32**. Fitness attained by a strong species, unchanging (stable) environment

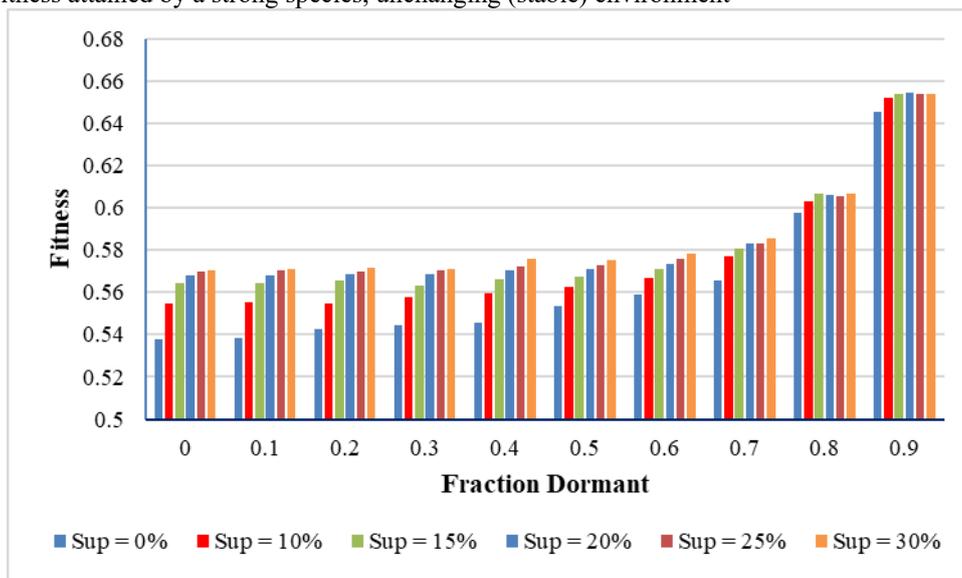

*Parameters*. $p_1$ = 0.50; $p_2$ = 0.90; $p_4$ = 0.01; $T$ = 100; 10,000 iterations; $v\_def$ = -0.15 for strong species; Dormancy fraction (*dorm_fr*) = 0 and 0.90 for Nil and Sporadic Dormancy respectively.

<u>Section B5</u>**.** Fitness gain (or loss) on account of dormancy, under (a) varying rate of migration of population-code traits to progeny **($p_1$)** and (b) varying rate of assimilation of genetic traits of fitter members into the population code **($p_2$)**

**fig. S33**. Gain in fitness due to sporadic dormancy, under varying rate of migration of population-code traits to progeny (***$p_1$***), for a weak species

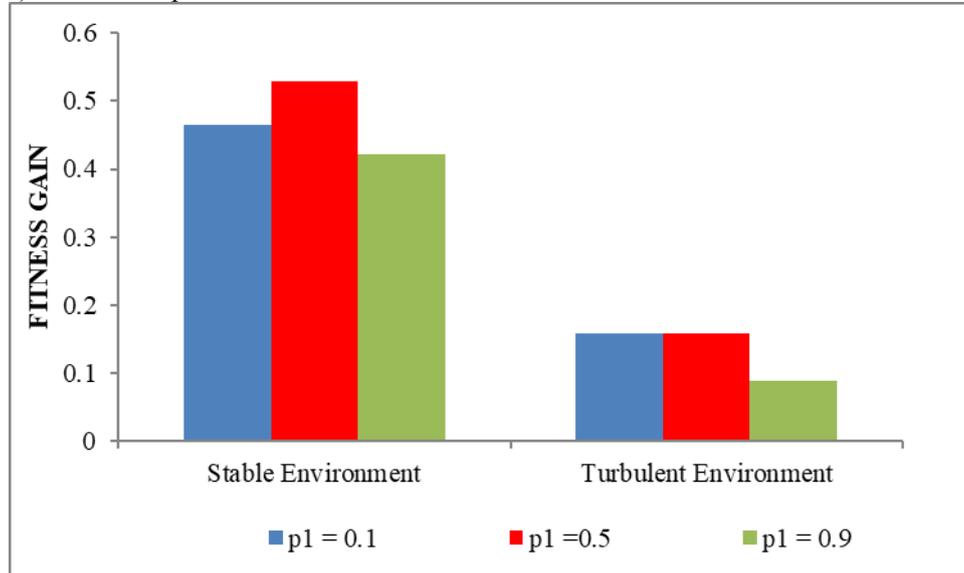

*Parameters.* $p_2$ = 0.50; $p_4$ = 0 for stable environment, 0.01 for turbulent environment; *T* = 100; *v_def* = 0.15 for weak species; 10,000 iterations; Fitness at Nil Dormancy is subtracted from Fitness at Dormancy Fraction (*dorm_fr*) = 0.90 to compute FITNESS GAIN.

**fig. S34**. Gain in fitness due to sporadic dormancy, under varying rate of assimilation of genetic traits of fitter members into the population code (***$p_2$***), for a weak species

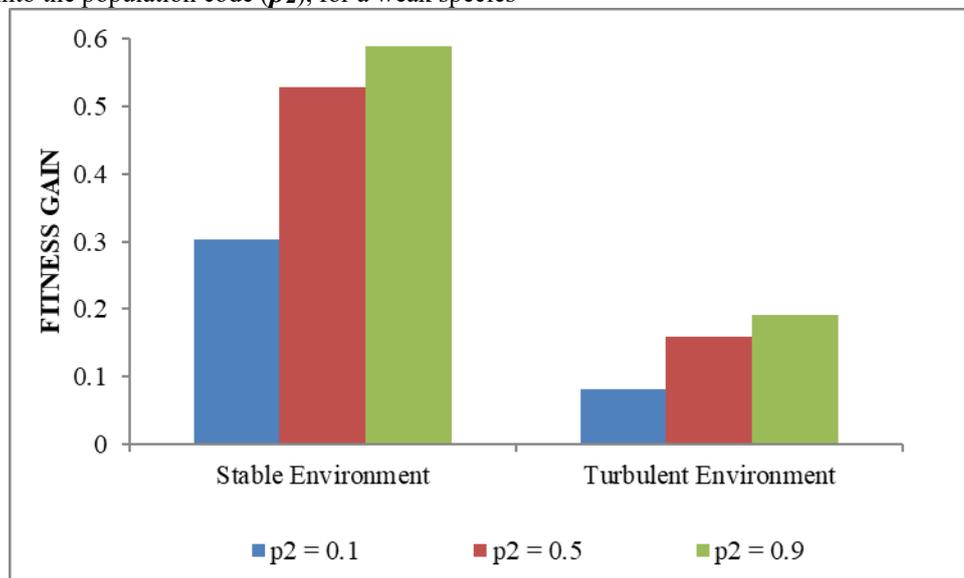

*Parameters.* $p_1$ = 0.50; $p_4$ = 0 for stable environment, 0.01 for turbulent environment; *T* = 100; *v_def* = 0.15 for weak species; 10,000 iterations; Fitness at Nil Dormancy is subtracted from Fitness at Dormancy Fraction (*dorm_fr*) = 0.90 to compute FITNESS GAIN.

**fig. S35**. Gain/Loss in fitness due to sporadic dormancy, under varying rate of migration of population-code traits to progeny (***p*$_1$**), for a strong species

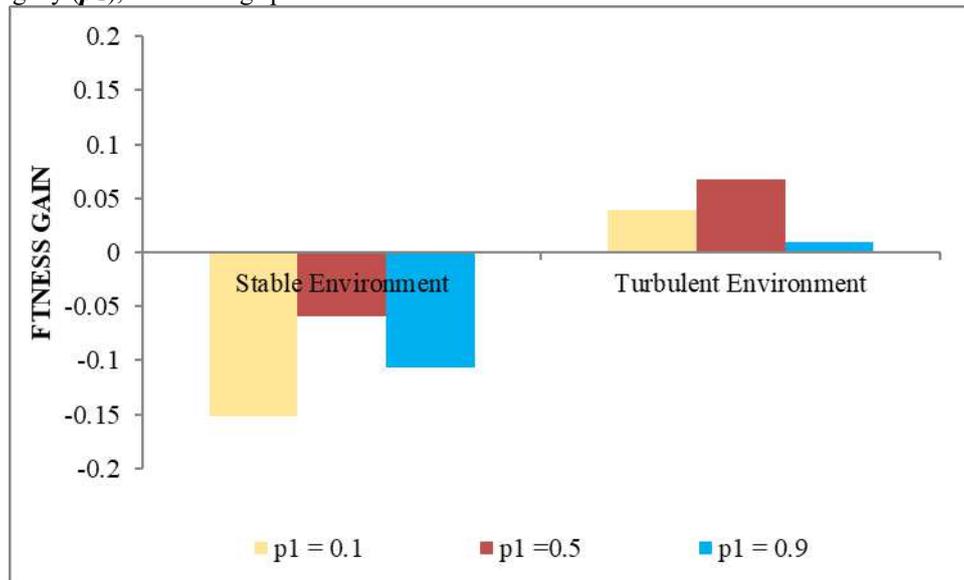

*Parameters*. ***p*$_2$** = 0.50; ***p*$_4$** = 0 for stable environment, 0.01 for turbulent environment; *T* = 100; *v_def* = -0.15 for strong species; 10,000 iterations; Fitness at Nil Dormancy is subtracted from Fitness at Dormancy Fraction (*dorm_fr*) = 0.90 to compute FITNESS GAIN (A negative value signifies FITNESS LOSS).

**fig. S36**. Gain/Loss in fitness due to sporadic dormancy, under varying rate of assimilation of genetic traits of fitter members into the population code (***p*$_2$**), for a strong species

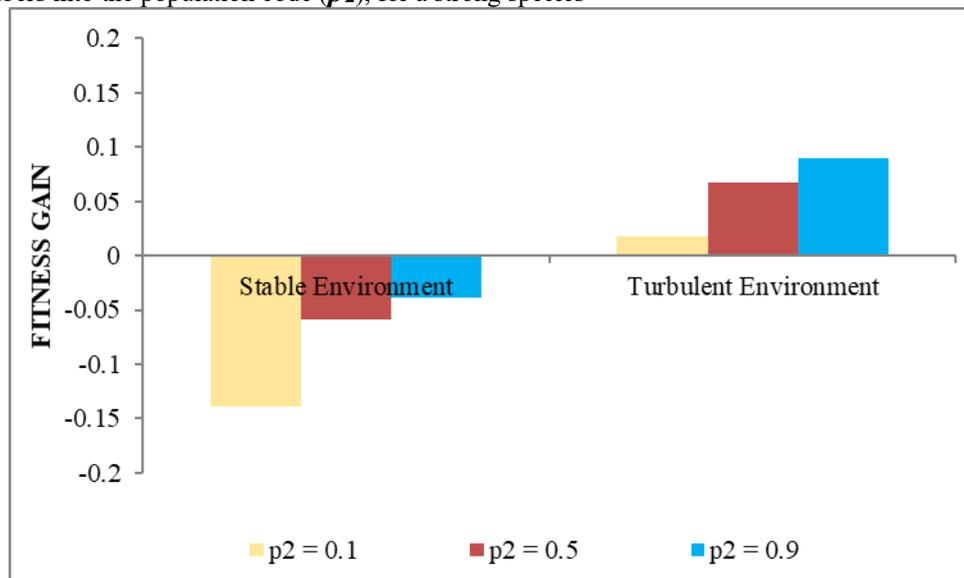

*Parameters*. ***p*$_1$** = 0.50; ***p*$_4$** = 0 for stable environment, 0.01 for turbulent environment; *T* = 100; *v_def* = -0.15 for strong species; 10,000 iterations; Fitness at Nil Dormancy is subtracted from Fitness at Dormancy Fraction (*dorm_fr*) = 0.90 to compute FITNESS GAIN (A negative value signifies FITNESS LOSS).